\documentclass[aps,prb,twocolumn,reprint,groupedaddress]{revtex4-2}
\usepackage{graphicx}
\usepackage{amsmath}
\usepackage{mathtools}
\usepackage{bbold}
\usepackage[usenames, dvipsnames]{color}
\usepackage{float}

\usepackage{hyperref}

\usepackage{xcolor}
\usepackage{cancel}
\usepackage{soul}
\usepackage{bm}
\usepackage[normalem]{ulem}
\usepackage{comment}

\allowdisplaybreaks

\begin{document}

\title{Spin modulations in the Rashba-Hubbard chain---a tensor network study}

\author{Jozef \textsc{Genzor}$^{1}$}
\email{jozef.genzor@gmail.com}
\author{Roman \textsc{Krčmár}$^{1}$}
\author{Andrej \textsc{Gendiar}$^{1}$}
\author{Chia-Min \textsc{Chung}$^{2, 3,4}$}
\author{Denis \textsc{Kochan}$^{1, 5}$}
\email{denis.kochan@savba.sk}

\affiliation{$^1$Institute of Physics, Slovak Academy of Sciences, SK-845 11, Bratislava, Slovakia}
\affiliation{$^2$Department of Electrophysics, National Yang Ming Chiao Tung University, Hsinchu 300, Taiwan}
\affiliation{$^3$Physics Division, National Center for Theoretical Sciences, Taipei 10617, Taiwan}
\affiliation{$^4$Center for Theoretical and Computational Physics, National Yang Ming Chiao Tung University, Hsinchu 300093, Taiwan}
\affiliation{$^5$Department of Physics and Center for Quantum Frontiers of Research and Technology (QFort), National Cheng Kung University, Tainan 70101, Taiwan}

\date{\today}

\begin{abstract}
Uniform spin-orbit coupling in an open single-band Hubbard chain is an exactly removable \(SU(2)\) gauge field at the Hamiltonian level, but not at the level of laboratory-frame spin correlations. We study this separation using density matrix renormalization group calculations for the repulsive one-dimensional Rashba-Hubbard chain. For open boundary conditions, a site-dependent spin rotation maps the model with hopping 
\(t\) and Rashba spin-orbit strength \(\lambda\) onto the ordinary Hubbard chain with renormalized hopping \(t_\lambda=\sqrt{t^2+\lambda^2}\). Consequently, charge and energy diagnostics are affected only through the bandwidth renormalization, which is quadratic in weak \(\lambda/t\). Spin correlations, however, respond already at linear order because the same transformation rotates the local spin basis by the wave vector \(k_{\rm so}=2\arctan(\lambda/t)\). We use DMRG to verify this observable consequence across the filling diagram of finite open chains. The filling structure follows the gauge-equivalent Hubbard model, whereas the spin structure factor shows the predicted spin-orbit sidebands. A dominant Hubbard-chain magnetic wave vector \(k_0\) is transformed into components at \(k_0\pm k_{\rm so}\), folded into the open-chain Brillouin zone. At half filling, where \(k_0=\pi\), the two sidebands fold onto a single, in-plane, spin spiral wave with \(k=\pi-k_{\rm so}<\pi\). Away from half filling, the incommensurate Hubbard spin response splits into two distinct spin-orbit-shifted components, producing a real-space beating pattern. Our results provide a filling-resolved tensor-network benchmark for the exactly removable limit of one-dimensional spin-orbit coupling, and establish a controlled reference point for ladders, multiorbital chains, rings, proximitized wires, and higher-dimensional Hubbard systems where spin-orbit coupling can no longer be gauged away.
\end{abstract}

\maketitle

\section{Introduction}

Spin--orbit coupling (SOC) is one of the standard routes by which spin becomes tied to motion in low-dimensional quantum matter. In one-dimensional systems this coupling underlies helical liquids, spin-selective transport, and the semiconductor-nanowire route to Majorana bound states, where SOC must act together with Zeeman splitting and superconducting proximity coupling~\cite{governale2004rashba,oreg2010helical,lutchyn2010majorana}. In parallel, ultracold atoms have made SOC a tunable synthetic gauge field rather than a fixed material parameter~\cite{lin2011spin,galitski2013spin,zhai2015degenerate}. These developments make one-dimensional SOC systems a natural meeting point of spintronics, strongly correlated physics, cold-atom simulation, and topological superconductivity.

There is, however, a basic distinction that is easily missed. SOC can be important in two very different ways. It can change the spectrum and generate new low-energy phases, as in helical or topological nanowire settings. Or it can be spectrally removable while remaining visible in spin-resolved observables. The single-band Hubbard chain with uniform Rashba-like SOC and open boundary conditions realizes the second, more subtle, possibility. It is therefore not a model where uniform SOC creates a new bulk phase. It is instead a controlled example in which the spectrum is ``gauge-trivial'', but the spin texture measured in the laboratory frame is ``gauge-visible''.

As discussed by Kaplan~\cite{kaplan1983single}, for an open chain, the uniform SOC hopping can be removed by a local spin rotation, which maps the SOC-coupled Hubbard model with hopping \(t\) and SOC strength \(\lambda\) onto an ordinary Hubbard chain with the renormalized hopping
\begin{equation}
    t_\lambda=\sqrt{t^2+\lambda^2}.
\end{equation}
Thus the SOC-coupled model at \((t,\lambda)\) is unitarily equivalent to the SOC-free Hubbard model at \((t_\lambda,0)\). Its many-body spectrum, charge gaps, and filling structure are those of the ordinary Hubbard chain after this bandwidth renormalization. For weak SOC, \(t_\lambda=t[1+\tfrac{1}{2}(\lambda/t)^2+\cdots]\), so the spectral and filling-boundary shifts are only second order in \(\lambda/t\). Goth and Assaad later developed this equivalence beyond the Hamiltonian level, and open boundary conditions deriving consequences for spectral functions, spin-resolved ARPES, QMC simulations, Heisenberg limit, and generalized Hubbard models~\cite{goth2014equivalence}. Persistent spin currents in the Rashba-Hubbard rings were studied by means of a nonperturbative treatment in combination with the Bethe ansatz by Fujimoto and Kawakami \cite{Fujimoto:PhysRevB.48.17406}.

The ordinary one-dimensional Hubbard model is already a highly constrained reference problem. It is exactly solvable by Bethe ansatz, displays a half-filled Mott regime for repulsive interactions, and realizes Luttinger-liquid physics away from half filling~\cite{lieb1968absence,essler2005hubbard,giamarchi2003quantum}. This makes the SOC-coupled chain useful not because it escapes the Hubbard paradigm, but because it tests how a gauge transformation acts on observables. Charge quantities and spectra follow the gauge-equivalent Hubbard chain. Spin observables do not. The local spin basis is rotated from site to site, so the laboratory operator \(S^z_j\) becomes a site-dependent mixture of transverse spin components in the SOC-free frame.

This immediately predicts the central spin-texture effect. If the SOC-free Hubbard chain has a dominant spin-correlations at wave-vector 
\(k_0\), the SOC-coupled chain acquires spin-correlation components shifted by
\begin{equation}
    k_0\pm k_{\rm so}= k_0 \pm 2\arctan(\lambda/t),
\end{equation}
modulo reciprocal-lattice vectors. At half filling, where the SOC-free spin correlations are staggered, the two shifted components are equivalent after Brillouin-zone folding, and the observed response is a single SOC-induced spiral modulation. Away from half filling, the SOC-free spin response is already incommensurate, and the same local rotation splits the original spin-correlation peak into two distinct SOC-dependent components. The effect is therefore not a new phase transition, but a precise and measurable displacement of spin-correlation weight.

This perspective also gives a direct experimental interpretation. In cold-atom Hubbard chains, spin- and density-resolved quantum gas microscopy can access real-space spin correlations and their Fourier components~\cite{boll2016spin,hilker2017revealing}. Synthetic SOC then provides a controlled way to tune the rotation angle and test the predicted shift of the spin-correlation peak. In solid-state quantum wires the same measurement is less direct, but SOC-induced spin textures can be probed through spin-resolved spectroscopy, spin-selective tunneling, or transport signatures of the spin-precession length; related spin--orbit density-wave textures have been observed in atomic wire systems~\cite{brand2015correlated}. The present model should therefore be viewed as a benchmark: before adding Zeeman fields, superconductivity, disorder, multiband structure, or interactions beyond the on-site Hubbard term, it gives the exact gauge baseline for what uniform SOC alone can and cannot do.

In this work we use density matrix renormalization group (DMRG) calculations to provide a filling-resolved real-space benchmark of this physics. We compute the filling, energy, entanglement entropy, boundary-pinned spin profiles, and spin--spin correlations of open chains up to \(L=224\). The filling diagram is used as a consistency check of the Kaplan site-resolved spin-rotation mapping \cite{kaplan1983single}: empty, partially filled, half-filled, and fully filled regimes follow the ordinary Hubbard-chain structure with the expected \(t\mapsto t_\lambda\) renormalization. The main result is the spin-correlation analysis. We show that the half-filled chain exhibits the predicted single folded spiral modulation, while the partially filled regimes display the expected splitting into two SOC-dependent spin-correlation components.

Our contribution is therefore not the rediscovery of the gauge mapping itself~\cite{kaplan1983single,Fujimoto:PhysRevB.48.17406,goth2014equivalence}. What we add is a systematic DMRG account of how the observable spin modulation evolves across the filling diagram in finite open chains. This provides a numerical reference point for more complex situations in which SOC is no longer a removable single-band gauge field, such as multiorbital chains~\cite{kaushal2017dmrg,li2016nonlocal}, two-dimensional Rashba--Hubbard systems~\cite{kennedy2022magnetism,PhysRevB.107.045123,PhysRevB.107.224427}, or proximitized quantum wires where SOC participates in helical~\cite{mendieta2026transport} and topological superconducting physics~\cite{oreg2010helical,lutchyn2010majorana}.

The paper is organized as follows. In Sec.~\ref{sec:model} we define the 1D Rashba-Hubbard model $=$ Hubbard model with nearest neighbor SOC, introduce the observables and summarize the local spin-rotation and particle--hole mappings used throughout the analysis. In Sec.~\ref{sec:results} we present DMRG implementation and diagnostic, the filling diagram and analyze the SOC-induced spin modulations at half filling and away from half filling. We summarize the implications and limitations in Sec.~\ref{sec:conclusions}. Appendix~\ref{appendix:bound} gives the outer filling boundaries, while Appendix~\ref{appendix:RPFnoSOC} collects the SOC-free real-space profiles used to extract the baseline modulation wavelengths.

\section{Rashba-Hubbard chain and its symmetries}\label{sec:model}

\subsection{Model Hamiltonian}

Consider a 1D chain of $L$ sites, where two fermions with opposite spins can sit on the same site.
The total Hamiltonian ${\cal H}$ for the 1D Hubbard model with spin-orbit coupling (SOC), boundary magnetic field, and on-site interaction \( U \) is written as
\begin{equation}
\label{eq:Hamiltonian}
    {\cal H}(t,\lambda,\mu,U) = H_0 + H_{\rm b}+ H_{\rm int} + H_{\rm soc},
\end{equation}
where
\begin{equation}
\begin{split}
    H_0 = & - t \sum\limits_{j=1}^{L-1} \sum_{\sigma=\uparrow,\downarrow} \left( c^\dagger_{j, \sigma} c^{~}_{j+1, \sigma} + c^\dagger_{j+1, \sigma} c^{~}_{j, \sigma} \right) \\
    & -\mu \sum\limits_{j=1}^{L} \sum_{\sigma=\uparrow,\downarrow} c^\dagger_{j, \sigma} c^{~}_{j, \sigma}
\end{split}
\end{equation}
describes the nearest-neighbor hopping parameterized by a hybridization $t$ and the chemical potential $\mu$ for the fermions---creation and annihilation operators \( c^\dagger_{a, \sigma} \) and \( c_{a, \sigma} \)---labeled by lattice site $a$ and spin projection~\mbox{$\sigma\in\{\uparrow,\downarrow\}\equiv\{+,-\}$}.
The hopping term $t$ is set to unity in the follow up DMRG calculations. A small boundary magnetic field $h\rightarrow 0$ entering $H_{\rm b}$ acts as a symmetry-breaking field along the $z$-direction
\begin{equation}
\begin{split}
H_{\rm b} = - & h\left(c^\dagger_{1,\uparrow}c_{1,\uparrow}-c^\dagger_{1,\downarrow}c_{1,\downarrow}\right) \\ 
+ & h\left(c^\dagger_{L,\uparrow}c_{L,\uparrow}-c^\dagger_{L,\downarrow}c_{L,\downarrow}\right) \,.
\end{split}
\label{eq:Hb}
\end{equation}
The repulsive electron-electron interaction parameterized by the Hubbard coupling strength $U>0$ reads
\begin{equation}
    H_{\rm int} = + U \sum\limits_{j=1}^{L} c^{\dagger}_{j, \uparrow}c^{~}_{j, \uparrow} c^{\dagger}_{j, \downarrow}c^{~}_{j, \downarrow}\,.
\end{equation}
Finally, the SOC introduces a spin-flip hopping $\lambda$ among the lattice sites
\begin{equation}
\begin{split}
H_{\rm soc,y} = & \ -\lambda \sum\limits_{j=1}^{L-1} \left( c^\dagger_{j, \uparrow} c^{~}_{j+1, \downarrow} - c^\dagger_{j+1, \uparrow} c^{~}_{j, \downarrow} \right) \\
& \ +\lambda \sum\limits_{j=1}^{L-1} \left( c^\dagger_{j, \downarrow} c^{~}_{j+1, \uparrow} - c^\dagger_{j+1, \downarrow} c^{~}_{j, \uparrow} \right)\,.
\end{split}
\label{eq:SOC}
\end{equation}
The present form derives from a tight-binding discretization of the conventional 1D Rashba SOC Hamiltonian 
\begin{equation}
H_{\rm soc,y} =\lambda k_x\sigma_y=
\frac{\hbar^2}{4m_e^2 c^2} (\boldsymbol{\nabla} V \times \mathbf{k}) \cdot \boldsymbol{\sigma}\,,
\end{equation}
assuming electrons are constrained to hop along $x$-axis, with a gradient field $\boldsymbol{\nabla} V$ pointing along $z$. Making a global unitary transformation in the spin space, or equivalently, changing the spin quantization axis from $z$ to $y$ axis, one can transform $H_{\rm soc,y} =\lambda k_x\sigma_y$ to $H_{\rm soc,z} =\lambda k_x\sigma_z$. Correspondingly, the SOC Hamiltonian takes the following tight-binding form
\begin{equation}
\begin{split}
H_{\rm soc,z} = & \ -i\lambda \sum\limits_{j=1}^{L-1} \left( c^\dagger_{j, \uparrow} c^{~}_{j+1, \uparrow} - c^\dagger_{j+1, \uparrow} c^{~}_{j, \uparrow} \right) \\
& \ +i\lambda \sum\limits_{j=1}^{L-1} \left( c^\dagger_{j, \downarrow} c^{~}_{j+1, \downarrow} - c^\dagger_{j+1, \downarrow} c^{~}_{j, \downarrow} \right)\,.
\end{split}
\label{eq:SOC}
\end{equation}
In what follows we use on-site particle number operator, $n_{\sigma, j}^{~}$, and spin operators, $S^{x,y,z}_j$, in units of $\hbar$, correspondingly for three Cartesian axes $x$, $y$, $z$, moreover we assume spin quantization along the $z$ axis, i.e.
\begin{equation}
n_{\sigma, j}^{~} 
= 
c_{\sigma, j}^\dagger c_{\sigma, j}^{~}
\ \ \text{and}\ \ 
S^{x,y,z}_j 
= 
\frac{1}{2} c_{\sigma, j}^\dagger\,(\sigma^{x,y,z})_{\sigma,\sigma'}^{~}\, c_{\sigma, j}^{~}\,,
\end{equation}
where $\sigma^{x,y,z}$ are conventional Pauli matrices.
The corresponding expectation values are computed in the quantum 
many-body ground state $| \psi_0 \rangle$ associated with the lowest energy $E_0$ of the Hamiltonian $\mathcal{H}$, Eq.~(\ref{eq:Hamiltonian}).

The spin-spin correlation function between $j$-th and $(j+\ell)$-th sites,
\begin{equation}
\langle S^z_{j} S^z_{j+\ell} \rangle = \langle \psi_0 | S^z_{j} S^z_{j+\ell} | \psi_0 \rangle,
\end{equation}
quantifies the spin alignment along the \( z \)-axis, providing us with insights into magnetic ordering and fluctuations in the system.
The filling factor $f$, $0\leq f \leq 1$, is defined as the number of particles per spin per site
\begin{equation}
f=\frac{N}{2L} \, ,
\end{equation}
where the expectation value gives the total number of particles
\begin{equation}
N = \sum\limits_{j=1}^{L} \sum\limits_{\sigma=\uparrow,\downarrow} \langle \psi_0 | n_{\sigma, j}^{~} | \psi_0 \rangle.
\end{equation}
In this convention, \( f = 1 \) corresponds to the full filling, i.e., there are $2L$ fermions sitting on the chain with the $L$ sites, while \( f = \tfrac{1}{2} \) corresponds to the half filling, i.e., $L$ fermions occupy the $L$-site chain on average. The energy per particle, \( \varepsilon \), is defined as the ground-state energy per measured number of particles
\begin{equation}
\varepsilon = \frac{E_0}{N} = \frac{\langle \psi_0 | {\cal H} | \psi_0 \rangle}{N} \, ,
\end{equation}
noticing that $\varepsilon\to0$ if $f\to0$.
The energy per particle \( \varepsilon \) provides a useful measure of the system's ground-state properties across different filling regimes.

\subsection{Symmetries of the Rashba-Hubbard chain}\label{sec:dualities}

When Hamiltonians at different coupling strengths $\bm g$ and $\bm g'$ are \emph{unitarily equivalent} up to a constant energy shift:
\begin{equation}
    H(\bm g)\ \xrightarrow{\ \mathcal{U}\ }\ H(\bm g')\equiv \mathcal{U}H(\bm g)\mathcal{U}^\dagger +\text{Const}(\bm g)\,,
\end{equation}
then the entire spectrum---more precisely all gaps, states $\psi_{\bm g}$ 
and so on---at coupling $\bm g$ get ``replicated'' at $\bm g'$. 
As a result, \emph{phase boundaries} are carried into \emph{phase boundaries}. Likewise, any expectation value (order parameter) defined by 
$m(\bm g)=\langle O\rangle_{\bm g}=\langle \psi_{\bm g}| O|\psi_{\bm g}\rangle$, transforms in the dual phase at $\bm g'$ according to
\begin{equation}
\label{Eq:OrderParameter}
    m(\bm g')
    =\langle \psi_{\bm g'}| O|\psi_{\bm g'}\rangle
    =\langle \psi_{\bm g}|\mathcal{U}^\dagger O\mathcal{U}|\psi_{\bm g}\rangle=
    \langle \mathcal{U}^\dagger O\mathcal{U}\rangle_{\bm g}.
\end{equation}

Here we show that the Hamiltonian for the Rashba-Hubbard chain, 
${\cal H}_{\bm g}={\cal H}(t,\lambda,\mu,U)$, with open boundary conditions and $h\rightarrow 0$, is in the above sense equivalent with the particle-hole reversed Rashba-Hubbard chain, ${\cal H}_{\bm g'}={\cal H}(t,\lambda,U-\mu,U)$, and also with an ordinary SOC-free Hubbard chain with
${\cal H}_{\bm g''}={\cal H}(\sqrt{t^2+\lambda^2},0,\mu,U)$. 

Employing SOC in the form of $H_{\rm soc,y}$ while simultaneously changing the spin quantization from $z$ to $y$ axis---we get the Hamiltonian for the Rashba-Hubbard chain in the form (to emphasize a change of spin quantization axis we use for spin the subscript $s$ instead of $\sigma$)
\begin{equation}
\begin{split}
{\cal H}(t,\lambda,\mu,U)=
&-\sum_{j=1}^{L-1}\sum_{s=\pm 1}
\Big[(t+i s\lambda)\,c^\dagger_{j,s}c_{j+1,s}+\text{h.c.}\Big]\\
&-\mu\sum_{j=1}^{L}\sum_{s=\pm 1} n_{j,s}
+U\sum_{j=1}^{L} n_{j,+}\,n_{j,-}\,,
\end{split}
\end{equation}
i.e. orbital and spin-orbital hybridizations combine into a complex spin-dependent hopping 
\begin{equation}
t+i s\lambda = t_\lambda\,e^{i s\theta_\lambda},
\ 
t_\lambda=\sqrt{t^2+\lambda^2},
\ 
\theta_\lambda=\arctan\!\frac{\lambda}{t}\,,   
\end{equation}
without loss of generality we assume $\theta_\lambda>0$.

\subsubsection{Particle-hole symmetry}

On a bipartite lattice (including a chain), the particle--hole transformation
\begin{equation}
\mathcal{P}:\quad c_{j,s}\mapsto (-1)^j c^\dagger_{j,s}
\quad\text{and}\quad
c^\dagger_{j,s}\mapsto (-1)^jc_{j,s}^{~}
\end{equation}
maps
\begin{equation}
\begin{split}
n_{j,s}=c^\dagger_{j,s}c_{j,s}\mapsto 1-n_{j,s},
\ \,
N=\sum_{j,s}n_{j,s}\mapsto 2L-N,
\end{split}
\end{equation}
and hence filling $f=\frac{N}{2L}\mapsto 1-f$. Applying $\mathcal{P}$ to the $\mu$ and $U$ terms gives
\begin{equation}
\begin{split}
-\mu\sum_{j,s}n_{j,s}
+U\sum_j n_{j,+}n_{j,-}
\mapsto-(U-\mu)\sum_{j,s}n_{j,s}
\\
+U\sum_j n_{j,+}n_{j,-}
+L(U-2\mu),    
\end{split}
\end{equation}
while the combined orbital and spin-orbital hopping term stays unchanged
\begin{equation}
    (t+i s\lambda)\,c^\dagger_{j,s}c_{j+1,s}^{~}
    \mapsto
    (t+i s\lambda)\,c^\dagger_{j,s}c_{j+1,s}^{~}\,.
\end{equation}
Therefore the full Hamiltonian transform as follows
\begin{equation}
\label{Eq:PHsymmetry}
\mathcal{P}\,\mathcal{H}(t,\lambda,\mu,U)\,\mathcal{P}^\dagger
=
\mathcal{H}(t,\lambda,U-\mu,U)+L(U-2\mu).
\end{equation}
The additive constant shifts all energies uniformly and does not affect eigenstates.
The invariant point is $\mu_{\rm inv}=U/2$. Thus the grand-canonical phase diagram is mirror-symmetric about the invariant point $\mu_{\rm inv}$.
Any phase boundary at $\mu=\mu_c$ has a partner boundary at $\mu=U-\mu_c$.
If the phase is self-dual under $\mathcal{P}$, the boundary can map onto itself only at $\mu=U/2$.

\subsubsection{``Gauging out'' spin-orbit coupling symmetry}

Defining the spin-dependent, lattice-site resolved, unitary transformation $\mathcal{U}_{\theta_\lambda}$ that acts on the ladder operators~\cite{kaplan1983single}:
\begin{equation}
\label{Eq:UnitaryUtheta}
\mathcal{U}_{\theta_\lambda}:
\quad
c_{j,s}^{~}=e^{-i s j\theta_\lambda}\,\tilde c_{j,s}^{~}
\quad\text{and}\quad
c^\dagger_{j,s}=e^{i s j\theta_\lambda}\,\tilde c_{j,s}^\dagger\,,
\end{equation}
the original complex hopping stratifies 
\begin{equation}
t_\lambda\,e^{i s\theta_\lambda}\,c^\dagger_{j,s}c_{j+1,s}^{~}
=
t_\lambda\,e^{i s(\theta_\lambda-\theta_\lambda)}\,
\tilde c^\dagger_{j,s}\tilde c_{j+1,s}^{~},
\end{equation}
yielding
\begin{equation}
H(t,\lambda,\mu,U)=
\mathcal{U}_{\theta_\lambda}^{\phantom{\dagger}}H(t_\lambda,0,\mu,U)\,\mathcal{U}_{\theta_\lambda}^\dagger\,.
\end{equation}
Therefore for open boundaries the SOC can be ``gauged away'' at the cost of renormalization of orbital hopping $t\mapsto t_\lambda=\sqrt{t^2+\lambda^2}$. Equivalently: the Rashba-Hubbard model at ${\bm g}=(t,\lambda)$ is spectrally dual to the SOC-free Hubbard model at ${\bm g'}=(t_\lambda,0)$.

On a ring, the hopping between the last and first lattice side would accumulate a global spin-dependent phase (twist/flux), 
$e^{i s L\theta_\lambda}$, so one cannot ``gauge-out'' SOC unless the $L\theta_\lambda=2\pi N$, what constraints allowed values of $t$, $\lambda$ and $L$. In what follows we focus only on the nanowires with open boundary conditions and hence we are not limited by any parameter constraints.

\subsubsection{Spin beating and spiral phase}

Unitary transformation $\mathcal{U}_{\theta_\lambda}$, Eq.~(\ref{Eq:UnitaryUtheta}), applies to creation and annihilation operators with spin projections along $y$ axis. 
Returning back to the original spin quantization axis along $z$ direction, the spin operators $S_j^{x,y,z}$ transform under $\mathcal{U}_{\theta_\lambda}$ as
\begin{equation}
\begin{aligned}
\mathcal{U}_{\theta_\lambda}^\dagger 
S_j^x\, 
\mathcal{U}_{\theta_\lambda}^{\phantom{\dagger}}
&=
\cos(2j\theta_\lambda)\, S_j^x
+
\sin(2j\theta_\lambda)\, S_j^z,
\\
\mathcal{U}_{\theta_\lambda}^\dagger 
S_j^y\, 
\mathcal{U}_{\theta_\lambda}^{\phantom{\dagger}}
&= S_j^y,
\\
\mathcal{U}_{\theta_\lambda}^\dagger 
S_j^z\, 
\mathcal{U}_{\theta_\lambda}^{\phantom{\dagger}}
&=
\cos(2j\theta_\lambda)\, S_j^z
-
\sin(2j\theta_\lambda)\, S_j^x .
\end{aligned}
\end{equation}
Therefore the expectation values in the Rashba-Hubbard model with 
${\bm g}=(t,\lambda)$ can be computed from the SOC-free Hubbard model at ${\bm g'}=(t_\lambda,0)$, 
see Eq.~(\ref{Eq:OrderParameter}), particularly for $S_j^x$ and $S_j^z$ it follows
\begin{equation}\label{Eq:SxSz}
\begin{aligned}
\langle S_j^x\rangle_{t,\lambda}
&=
\cos(2j\theta_\lambda)\,\langle S_j^x\rangle_{t_\lambda,0}
+
\sin(2j\theta_\lambda)\,\langle S_j^z\rangle_{t_\lambda,0}\,,
\\
\langle S_j^z\rangle_{t,\lambda}
&=
\cos(2j\theta_\lambda)\,\langle S_j^z\rangle_{t_\lambda,0}
-
\sin(2j\theta_\lambda)\,\langle S_j^x\rangle_{t_\lambda,0}\,.
\end{aligned}
\end{equation}

When the SOC-free Hubbard chain already possesses a spin modulation with a wave-vector $k_0$ (in units of inverse lattice constant), e.g.
\begin{equation}
\langle S_j^x\rangle_{t_\lambda,0}=A_x\cos(k_0 j),
\quad
\langle S_j^z\rangle_{t_\lambda,0}=A_z\cos(k_0 j),
\end{equation}
then using basic trigonometry 
\begin{equation*}
\begin{aligned}
\cos(2j\theta)\cos(k_0 j)
&=
\frac12\Big[\cos(k_0+2\theta)j+\cos(k_0-2\theta)j\Big],
\\
\sin(2j\theta)\cos(k_0 j)
&=
\frac12\Big[\sin(k_0+2\theta)j+\sin(2\theta-k_0)j\Big],
\end{aligned}
\end{equation*}
one sees that the spatial pattern of $\langle S_j^{x,z}\rangle_{t,\lambda}$ in the Rashba-Hubbard case generically acquires two beating components at shifted wave-vectors $k_{\lambda,\pm}$:
\begin{equation}
k_0 \;\longrightarrow\; k_{\lambda,\pm}=k_0\pm k_{\rm so}=k_0\pm 2\theta_\lambda \quad (\mathrm{mod}\;2\pi).
\end{equation}
So the generic effect of turning on the SOC in the Hubbard chain is a \emph{splitting and shifting} of the wave-vector of the dominant spin modulation, $k_0$, into two dominant modulations at $k_{\lambda,\pm}$, the splitting is proportional to $\theta_\lambda=\arctan{(\lambda/t)}$, independently of filling and strength of Hubbard $U$. For small $\lambda$
spin modulations develop already in a linear order in $\lambda/t\approx\arctan{(\lambda/t)}$, while the spectral ones scale quadratically $t_\lambda\approx t[1+\tfrac{1}{2}(\lambda/t)^2]$.
As a remark, for systems with the open boundary conditions as in our case, what matters are standing waves, therefore the wave-vectors $k$ and $-k$ in the first Brillouine zone are effectively equivalent, and therefore what plays a role is $k\in[0,\pi]$.\\

A special situation occurs for the Rashba--Hubbard chain at half filling in the large-\(U\) limit. 
In the SOC-free Hubbard chain, the half-filled state has a \emph{staggered}, antiferromagnetic (AFM), spin profile \(\langle S_j^{z}\rangle_{t_\lambda,0}\) with wave-vector \(k_0=\pi\), while its transverse spin components 
\(\langle S_j^{x}\rangle_{t_\lambda,0}\) and 
\(\langle S_j^{y}\rangle_{t_\lambda,0}\) vanish. 
According to Eqs.~(\ref{Eq:SxSz}), the corresponding spin profile in the Rashba--Hubbard chain acquires \emph{spiral spin modulation} with \(\langle S_j^{x,z}\rangle_{t_\lambda,0}\neq 0\):
\begin{equation}
\label{Eq:SxSz2}
\begin{aligned}
\langle S_j^x\rangle_{t,\lambda}
&=
\sin(2j\theta_\lambda)\,\langle S_j^z\rangle_{t_\lambda,0}\,,
\\
\langle S_j^z\rangle_{t,\lambda}
&=
\cos(2j\theta_\lambda)\,\langle S_j^z\rangle_{t_\lambda,0}\,.
\end{aligned}
\end{equation}
Moreover, $x$ and $z$ components are half-phase-shifted, and both develop the same spatial modulations at seemingly two dominant wave-vectors
\begin{equation}
    k_{\lambda,+}=\pi+2\theta_\lambda,
    \qquad
    k_{\lambda,-}=\pi-2\theta_\lambda .
\end{equation}
Since, the wave-vector \(k_{\lambda,+}=\pi+2\theta_\lambda\) lies outside the first Brillouin zone we shift it by a reciprocal lattice vector \(2\pi\) getting 
\(k_{\lambda,+}-2\pi=-(\pi-2\theta_\lambda)=-k_{\lambda,-}\). As the wave-vectors \(k\) and \(-k\) describe the same standing-wave modulation, two 
vectors, $k_{\lambda,+}$ and $k_{\lambda,-}$, are in-fact not independent at half filling and 
both reduce to a single folded wave-vector 
\(k_{\lambda,-}=\pi-2\theta_\lambda\in[0,\pi]\). 
Interestingly, for a generic $\lambda$, the wave-vector $k_{\lambda,-}$ would not be lattice-commensurate, i.e. $k_{\lambda,-}\neq \tfrac{p}{q}\tfrac{2\pi}{L}$ with $\tfrac{p}{q}$ rational.
%
%
So in the half-filling the Hubbard (H) and Rashba-Hubbard (RH) chains dominantly develop single spin modulations with the following characteristics:
\begin{align}
\label{Eq:HalfFillingKvectors}
\text{H:}\ &\text{commensurate AFM-phase}
&
k_0&=\pi,\nonumber
\\
&{}
\\
\text{RH:}\ &\text{incommensurate $xz$-spiral-phase}
&
k_{\lambda,-} &=\pi-2\theta_\lambda\,.\nonumber
\end{align}
Out of the half-filling the Rashba-Hubbard chain always develops a beating spin pattern composed of two dominant $k_{\lambda,+}$ and $k_{\lambda,-}$.

\section{Numerical Results}\label{sec:results}

\subsection{DMRG implementation}

We obtain the ground state $|\psi_0\rangle$ of the Hamiltonian ${\cal H}$, Eq.~(\ref{eq:Hamiltonian}), using the density-matrix renormalization group (DMRG)~\cite{white1992density,white1993algorithms} in its modern matrix-product-state (MPS) formulation~\cite{schollwock2011mps}, as implemented in the ITensor library~\cite{fishman2022itensor}.
An MPS represents the many-body wavefunction as a one-dimensional chain of site tensors linked by auxiliary (bond) indices, whose dimension --- the bond dimension $D$ --- controls how much entanglement the state can encode. Because ground states of one-dimensional systems are only weakly entangled (they obey an entanglement area law, up to logarithmic corrections at criticality), they are accurately represented by an MPS of moderate $D$, which makes DMRG a controlled and essentially numerically exact method for one-dimensional chains.
The ground state is found variationally, by minimizing the energy $E_0=\langle\psi_0|{\cal H}|\psi_0\rangle$ through repeated sweeps along the chain in which the tensors are optimized locally; the bond dimension $D$ bounds the number of retained Schmidt values and hence the truncation error. The method thus goes beyond mean-field approximations --- it retains the quantum correlations and entanglement of the interacting ($U\neq 0$) chain --- and yields real-space, finite-size information (density and spin profiles, correlation functions) that is not directly accessible from the Bethe-ansatz solution of the translationally invariant model.
All calculations are performed on open chains with the hopping fixed to $t=1$, for system sizes from $L=100$ up to $L=224$ and bond dimensions from $D=100$ up to $D=200$. The real-space spin profiles are obtained by applying a small anti-parallel boundary (pinning) field $h$ along the $z$ direction at the two ends of the chain [the $H_{\rm b}$ term in ${\cal H}$], which lifts the degeneracy of the symmetry-broken order ($h=0.1$ in practice); the spin--spin correlations are instead evaluated without any field ($h\rightarrow 0$). When convenient, the total particle number is fixed to the target filling directly in the DMRG sweeps. We verified that the results are converged in both $L$ and $D$ --- for instance, the dominant wavelengths for $L=200$, $D=200$ coincide with those for $L=102$, $D=100$ --- so that no finite-size scaling was necessary.
From the converged ground state we evaluate the total particle number $N$ and energy $E_0$, the local spin profiles $\langle S^z_j\rangle$ and $\langle S^x_j\rangle$, the spin--spin correlations $\langle S^z_j S^z_{j+\ell}\rangle$ together with their Fourier transform (used to extract the dominant wave-vectors), and the bipartite entanglement entropy described below.
The entanglement entropy is computed at the central bond by a singular value decomposition (SVD) of the MPS representation of $|\psi_0\rangle$. The MPS is first brought into canonical form, so that its orthogonality center sits at the central bond; the corresponding tensor $\psi$ is reshaped into a matrix and decomposed as
\begin{equation}
\psi_{xy} = \sum\limits_\alpha U_{x\alpha}^{~} s^{~}_\alpha V_{\alpha y}^\dagger .
\end{equation}
The singular values $s_\alpha$ (the Schmidt coefficients) encode the entanglement between the two halves of the chain and define the von Neumann entanglement entropy
\begin{equation}
S_{\text{E}} = - \sum_{\alpha} s_{\alpha}^2 \log s_{\alpha}^2 ,
\end{equation}
where $p_\alpha=s_\alpha^2$ are the eigenvalues of the reduced density matrix of one half. A large $S_{\text{E}}$ signals strong entanglement between the two halves, a small value weak entanglement.

\subsection{Phase diagram}

\begin{figure}
\includegraphics[width=0.48\textwidth]{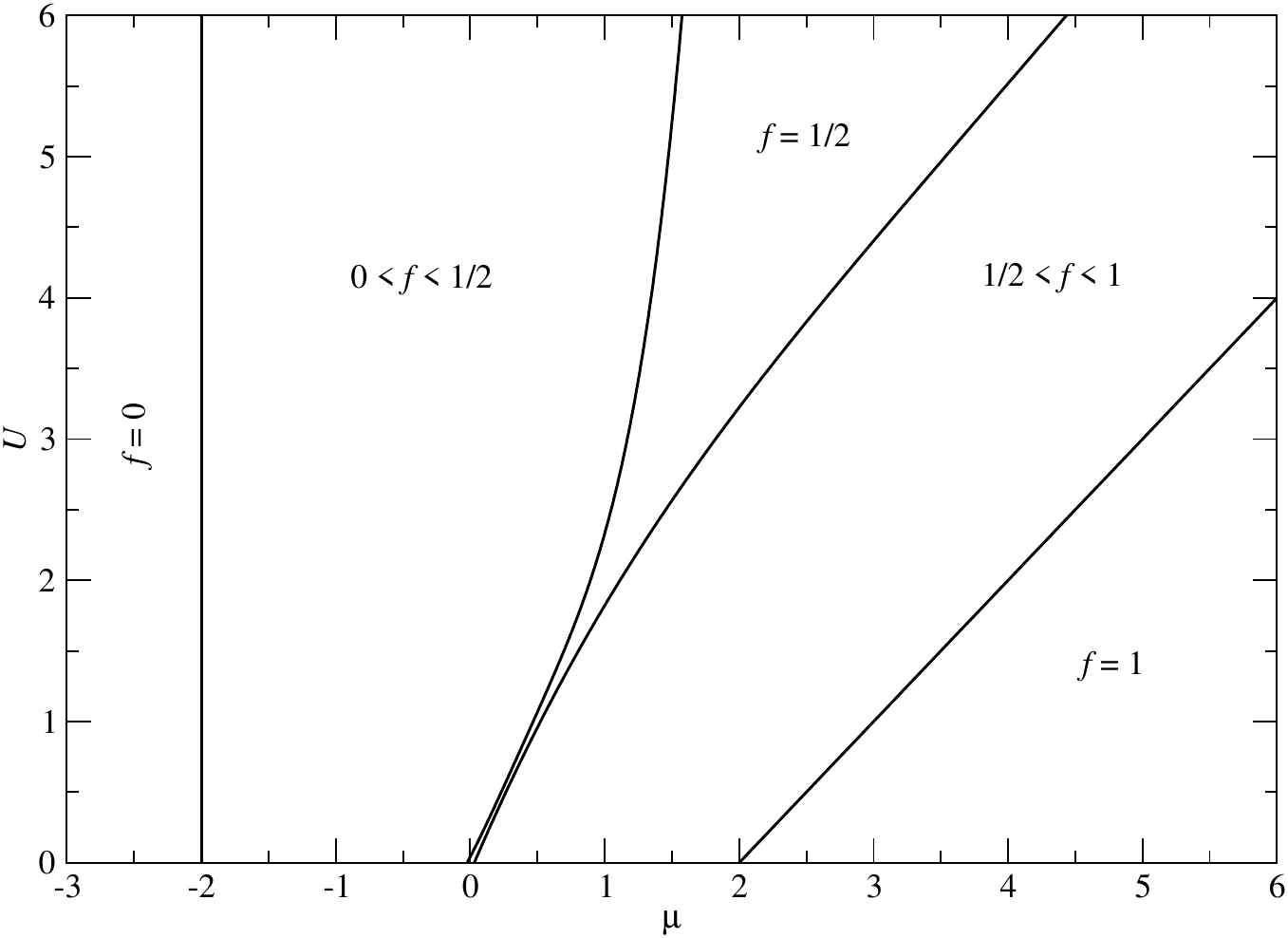}
\caption{
Filling regimes of the open one-dimensional Hubbard chain without spin-orbit coupling. The regions are classified by the filling factor $f=N/(2L)$: empty chain $(f=0)$, partially filled below half filling $(0<f<1/2)$, half-filled Mott plateau $(f=1/2)$, partially filled above half filling $(1/2<f<1)$, and full chain $(f=1)$. The calculation was performed for $L=100$ and bond dimension $D=100$. The roughness of the plateau boundaries reflects the finite sampling step in $(\mu,U)$, not an additional physical structure. 
}
\label{fig:phase_diagram_lambda_0}
\end{figure}

\begin{figure}
\includegraphics[width=0.48\textwidth]{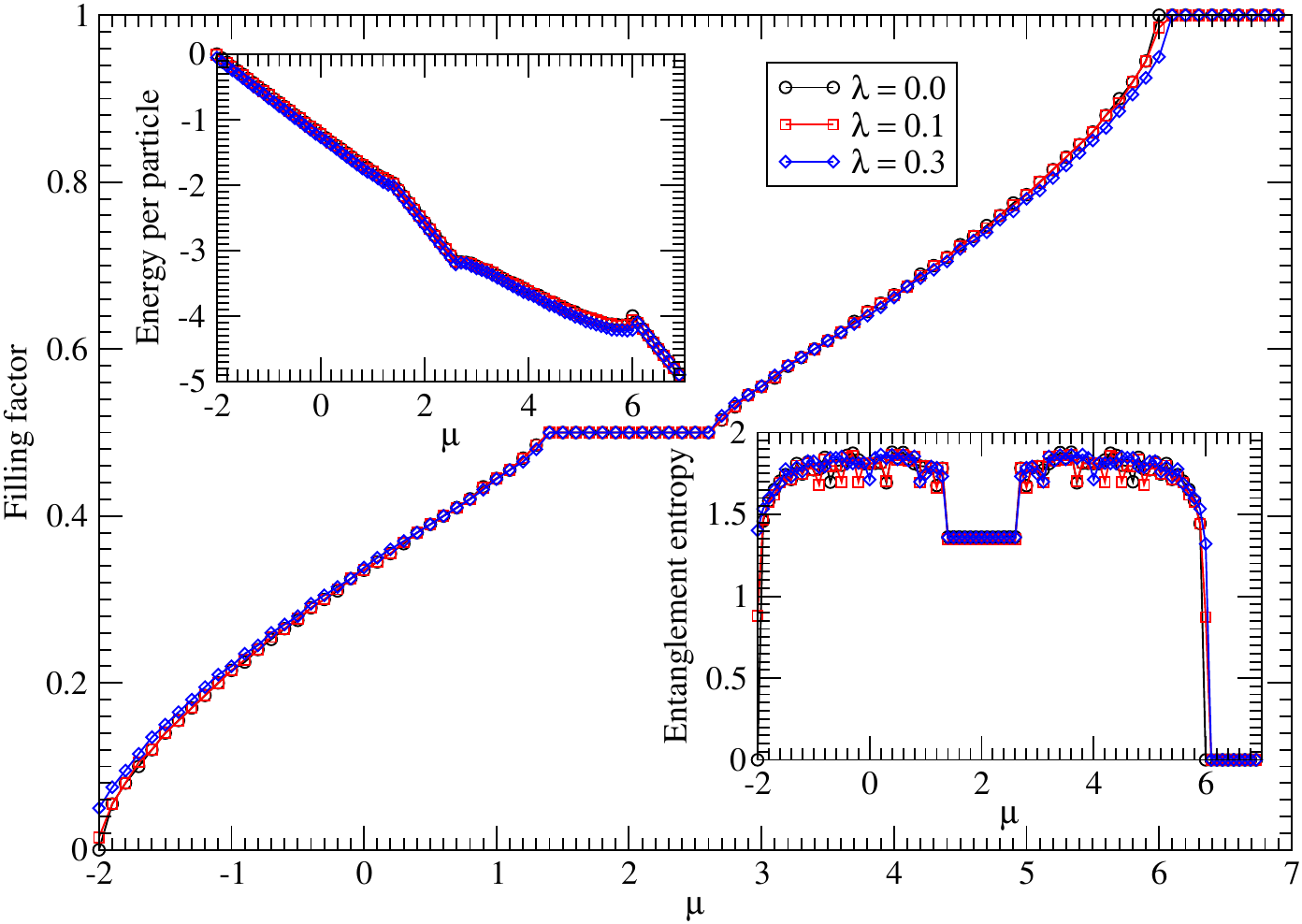}
\caption{
One-dimensional cut through the filling diagram at $U=4$, comparing $\lambda=0$, $\lambda=0.1$, and $\lambda=0.3$. The main panel shows the filling factor $f=N/(2L)$ as a function of chemical potential $\mu$. The upper-left inset shows the ground-state energy per particle, and the lower-right inset shows the bipartite entanglement entropy at the central bond. The weak dependence of these charge and energy diagnostics on $\lambda$ is consistent with the open-chain mapping $t\mapsto t_\lambda=\sqrt{t^2+\lambda^2}$, which modifies the bandwidth only at order $\lambda^2/t$ for weak SOC. All data were obtained for $L=100$, $D=100$, and $t=1$.
}
\label{fig:filling_entanglement_energy_U4_N100}
\end{figure}

We have calculated the phase diagram for the 1D Hubbard model without SOC for $\mu \in \left(-3, 6\right)$ and $U \in \left(0, 6\right)$, see Fig.~\ref{fig:phase_diagram_lambda_0}.  
Based on the occupancy $f=N/(2L)$, we identified five different phases: empty filling ($f = 0$), under-half filling ($f < 0.5$), half filling 
($f = 0.5$), above-half filling ($0.5 < f < 1$), and full filling 
($f = 1$). 
The roughness of the boundaries of the half-filling region is just an artifact of the finite scanning step used in sampling the data points in Fig.~\ref{fig:phase_diagram_lambda_0}. 
The phase diagram does not change significantly when including a small SOC parameter $\lambda$.
For example, the boundaries between the different phases in Fig.~\ref{fig:phase_diagram_lambda_0} would shift by $\approx 0.1$ in $\mu$ when including $\lambda = 0.1$ (not presented here).
To see how inclusion of the SOC parameter $\lambda$ influences the phase diagram, we display in Fig.~\ref{fig:filling_entanglement_energy_U4_N100} a cut at $U=4$.
For $\lambda=0$, the boundary between the empty filling and the under-half filling is at $\mu = -2$, see Appendix~\ref{appendix:bound}.
The boundary between the under-half filling and the half-filling region is at $\mu \approx 1.4$, whereas the boundary between the half-filling and the above-half filling lies at $\mu \approx 2.6$.
And finally, the boundary between the above-half filling and the full filling lies at $\mu = 6$, see Appendix~\ref{appendix:bound}.
These boundaries can be determined independently by looking at the filling factor, energy per particle, and entanglement entropy, see Fig.~\ref{fig:filling_entanglement_energy_U4_N100}.
We also compared these three quantities for $\lambda = \{0, 0.1, 0.3\}$, and one can see that the differences in the boundaries are really small.

\subsection{Half filling}

\begin{figure}
\includegraphics[width=0.48\textwidth]{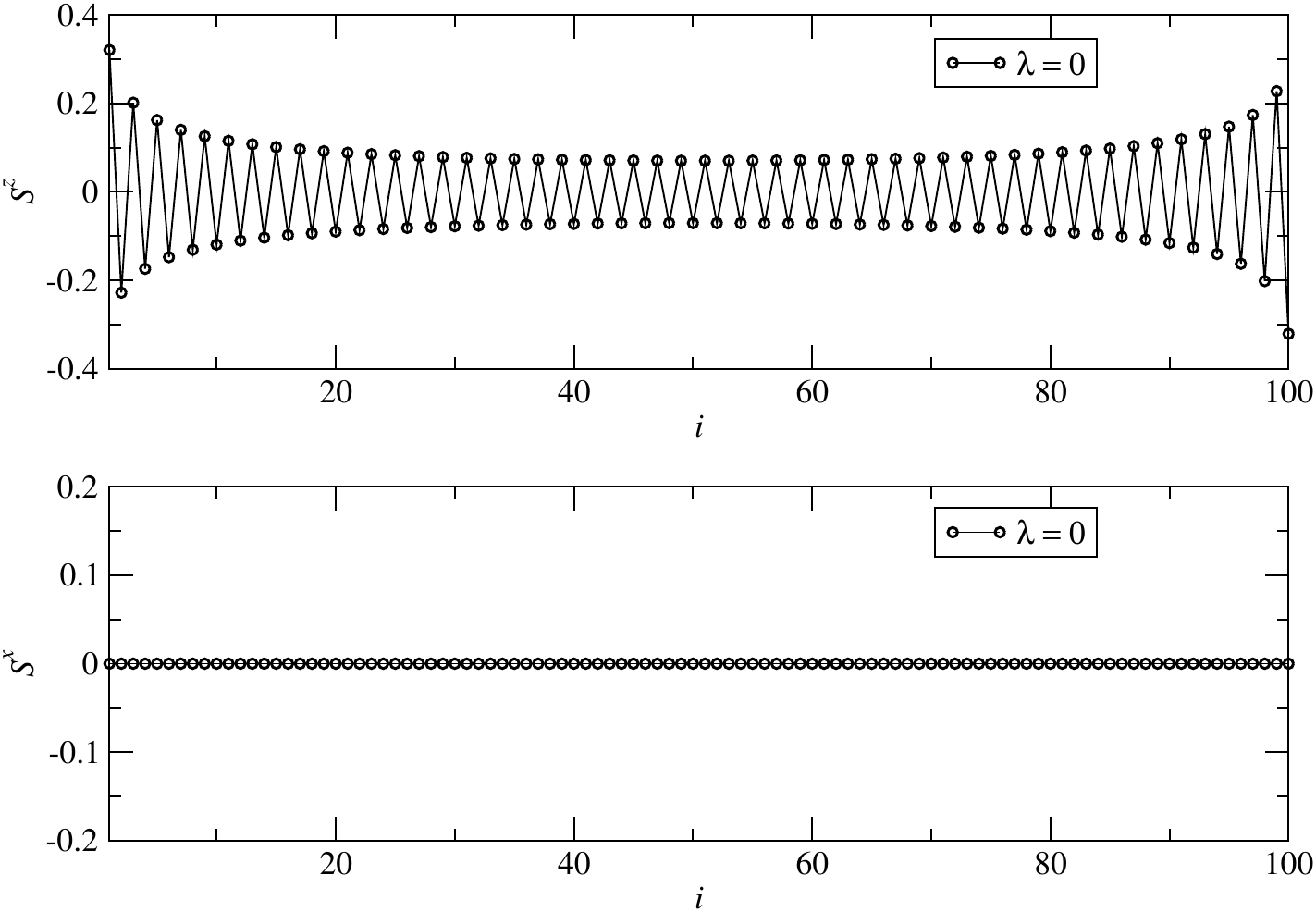}
\caption{
%
Boundary-pinned spin profile at half filling in the absence of spin-orbit coupling. The upper panel shows $\langle S^z_j\rangle$, while the lower panel shows $\langle S^x_j\rangle$. A weak antiparallel boundary field $h=0.1$ pins the staggered antiferromagnetic correlations in the $z$ direction, producing a real-space profile with dominant wave vector $k_0=\pi$. The transverse component remains zero within numerical accuracy. Parameters: $\lambda=0$, $\mu=3$, $U=10$, $L=100$, and $D=100$.
}
\label{fig:hf_mu3_U10_lam0.0}
\end{figure}

\begin{figure}
\includegraphics[width=0.48\textwidth]{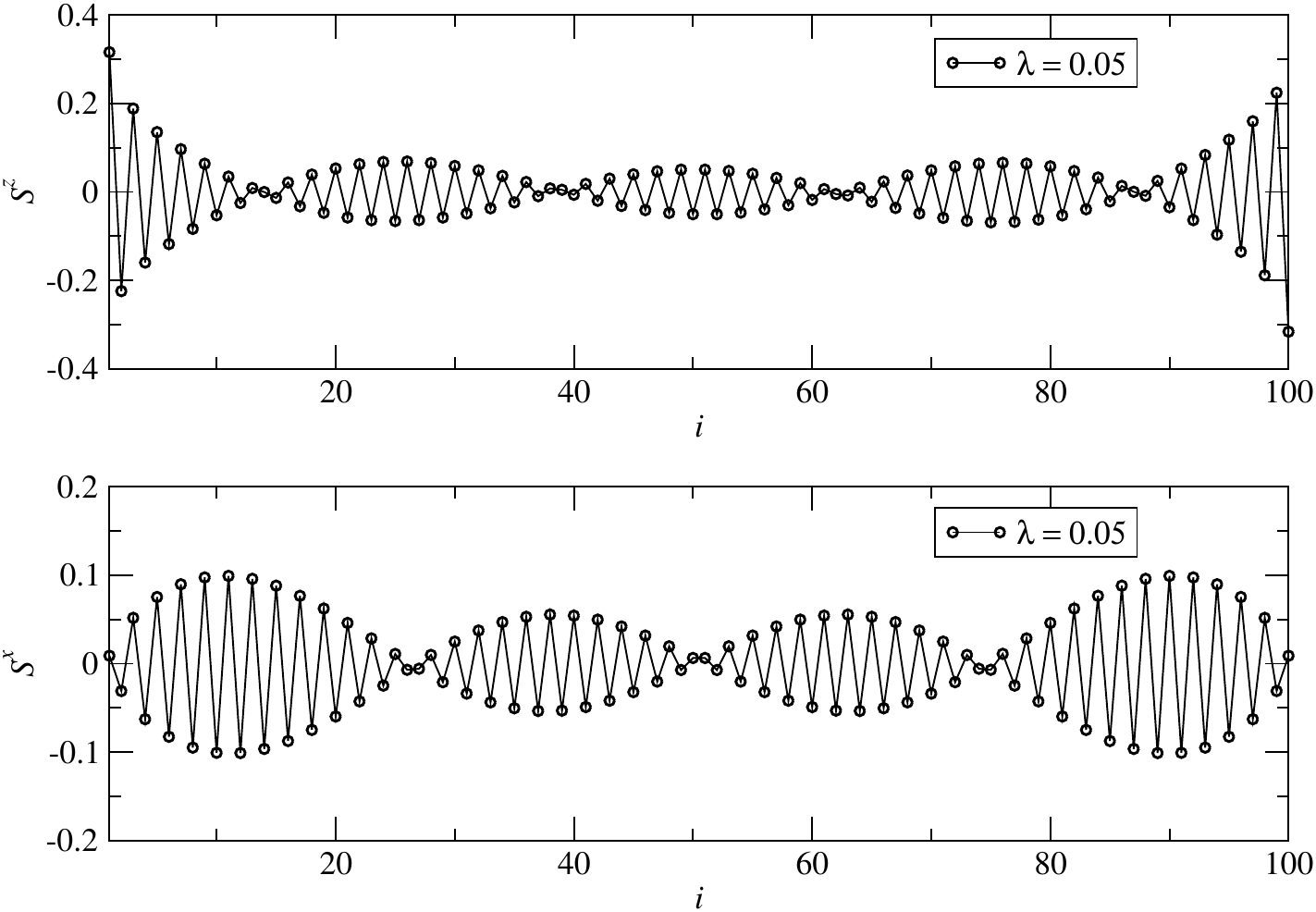}
\caption{
%
%
Boundary-pinned spin profile at half filling for weak spin-orbit coupling $\lambda=0.05$. The staggered $S^z$ profile of the ordinary Hubbard chain is locally rotated by the SOC wave vector 
$k_{\rm so}=2\arctan(\lambda/t)$, generating a finite $S^x$ component and a spiral texture in the $xz$ plane. The two components are shifted in phase, consistent with the local spin-rotation relation between the Rashba-Hubbard chain and the SOC-free Hubbard chain at $t_\lambda=\sqrt{t^2+\lambda^2}$. Parameters: $\mu=3$, $U=10$, $L=100$, $D=100$, $h=0.1$, and $t=1$.
}
\label{fig:hf_mu3_U10_lam0.05}
\end{figure}

\begin{figure}
\includegraphics[width=0.48\textwidth]{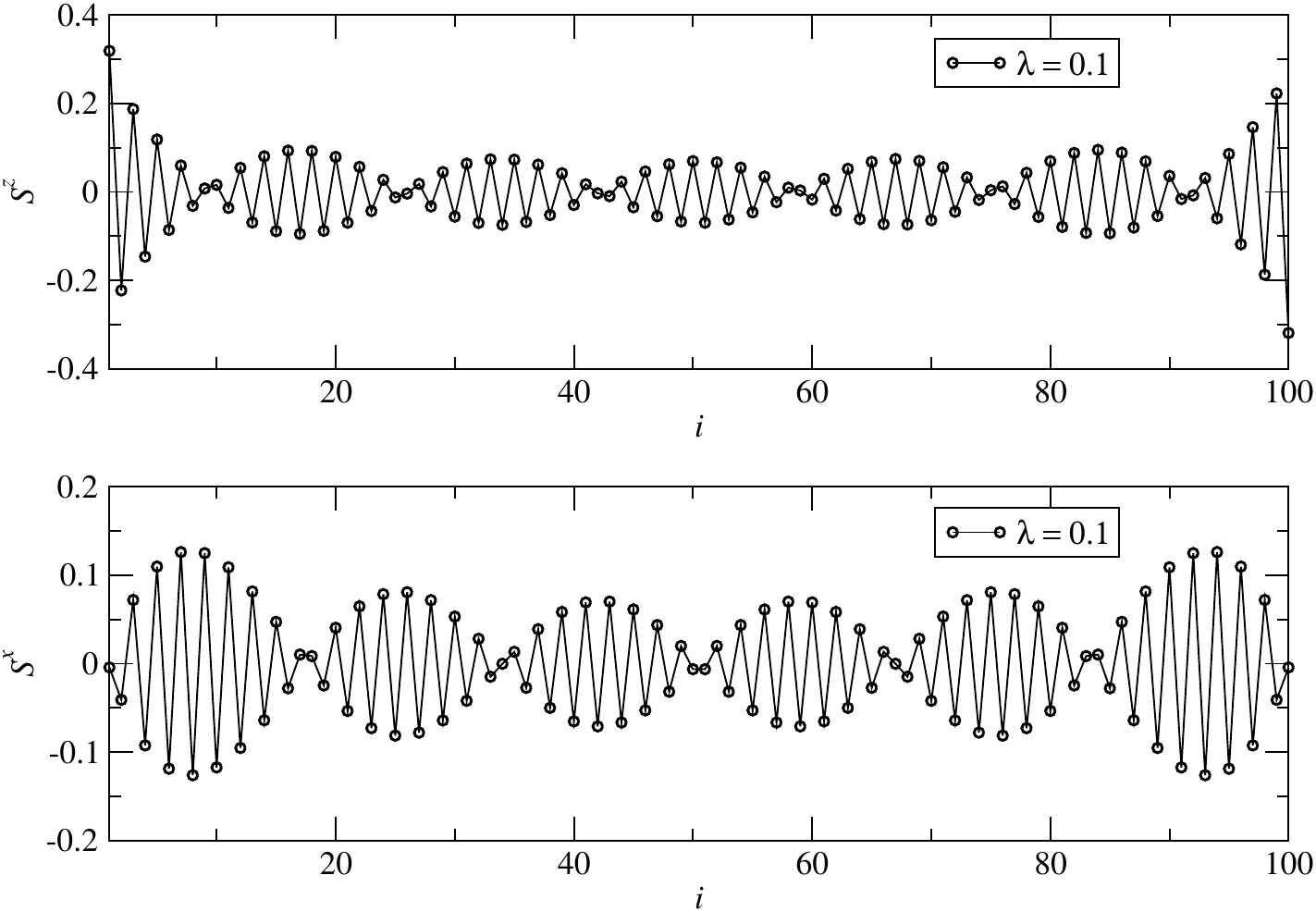}
\caption{
%
Boundary-pinned spin profile at half filling for $\lambda=0.1$. Compared with Fig.~\ref{fig:hf_mu3_U10_lam0.05}, the larger SOC increases the rotation wave vector $k_{\rm so}=2\arctan(\lambda/t)$, shortening the spiral pitch of the $xz$-plane spin texture. The result illustrates the linear-in-$\lambda$ response of spin observables, in contrast to the quadratic-in-$\lambda$ bandwidth renormalization controlling charge and energy diagnostics. Parameters: $\mu=3$, $U=10$, $L=100$, $D=100$, $h=0.1$, and $t=1$.
}
\label{fig:hf_mu3_U10_lam0.1}
\end{figure}

\begin{figure}
\includegraphics[width=0.48\textwidth]{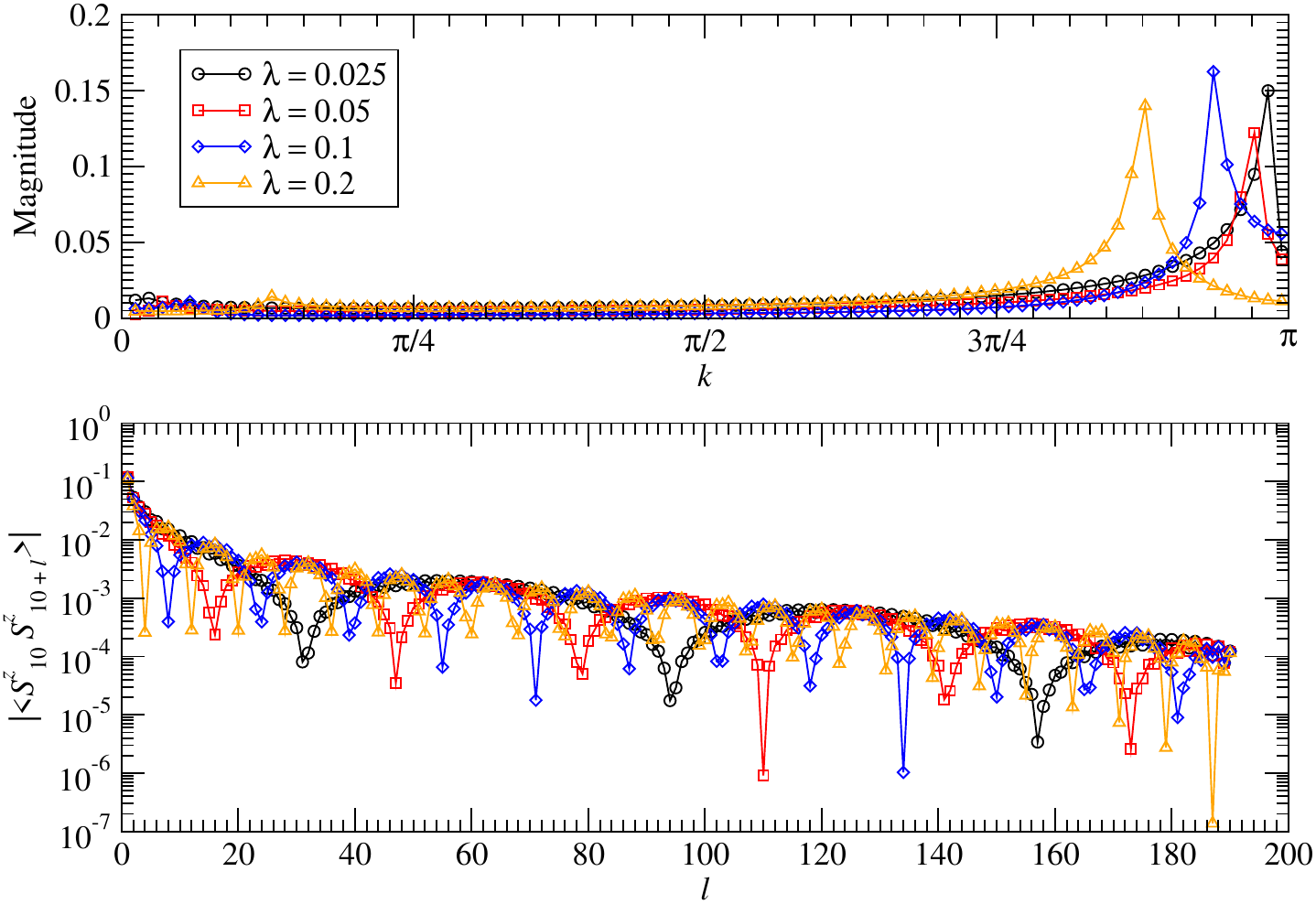}
\caption{
Field-free spin-spin correlations at half filling and their Fourier spectra for several SOC strengths. The real-space data show $|\langle S^z_{10}S^z_{10+\ell}\rangle|$, while the Fourier spectra identify the dominant modulation wave vector. At half filling, the two SOC-shifted sidebands $\pi\pm k_{\rm so}$, with $k_{\rm so}=2\arctan(\lambda/t)$, are equivalent after Brillouin-zone and open-chain folding, leaving a single folded component at $k=\pi-k_{\rm so}$. The corresponding wavelengths are approximately $2.0357$, $2.0602$, $2.1375$, and $2.2800$ for $\lambda=0.025$, $\lambda=0.05$, $\lambda=0.1$, and $\lambda=0.2$, respectively. Parameters: $\mu=3$, $U=10$, $L=200$, $D=100$, and $h=0$.
}
\label{fig:corr_mi3_U10_dim100}
\end{figure}

\begin{figure}
\includegraphics[width=0.48\textwidth]{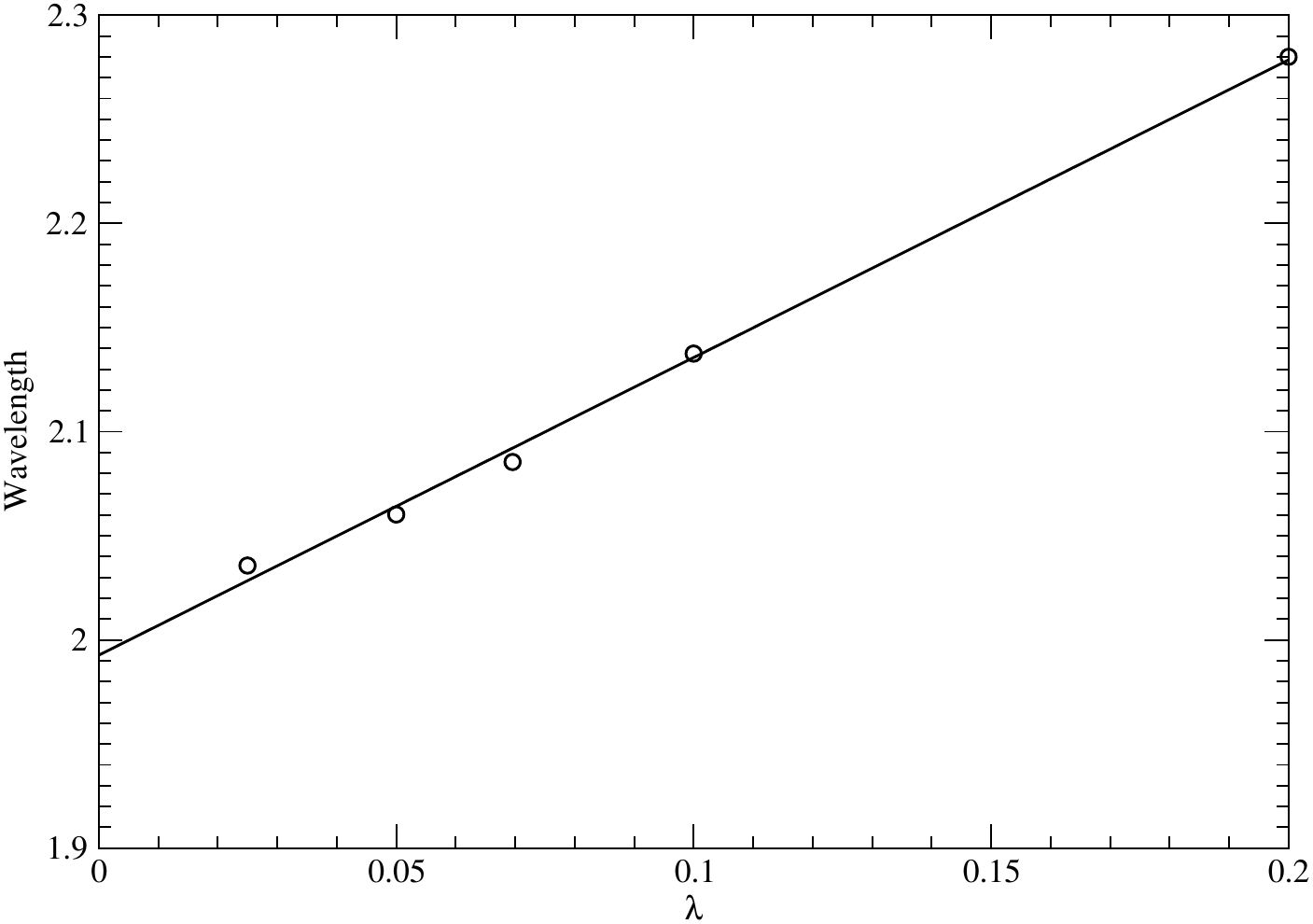}
\caption{
%
Dominant half-filled spin-modulation wavelength as a function of SOC strength. The wavelength extracted from the Fourier spectra follows the folded-sideband prediction $\ell(\lambda)=2\pi/[\pi-2\arctan(\lambda/t)]$ and approaches the staggered Hubbard-chain value $\ell=2$ as $\lambda\to0$. The approximately linear increase at weak SOC reflects 
$k_{\rm so}=2\arctan(\lambda/t)\simeq2\lambda/t$. Parameters: $\mu=3$, $U=10$, and $t=1$.
}
\label{fig:dft_period_vs_lambda_mu3_U10_paper}
\end{figure}

In the half-filling region without SOC, there is a long-range anti-ferromagnetic (AFM) ordering in the $z$-direction while there is no magnetization in the $x$-direction, see Fig.~\ref{fig:hf_mu3_U10_lam0.0}. 
To observe the AFM ordering in $S^z$, we applied an anti-parallel magnetic field in the $z$-direction at the boundaries only. 
However, when $\lambda > 0$, modulation of the $S^z$ expectation profile appears, which seems to be dependent on the value of $\lambda$, see Fig.~\ref{fig:hf_mu3_U10_lam0.05} for $\lambda = 0.05$ and compare to Fig.~\ref{fig:hf_mu3_U10_lam0.1} for $\lambda = 0.1$.
Moreover, when $\lambda > 0$, we also observe non-zero $S^x$ expectation, which exhibits a modulation similar to the case of $S^z$ but with a constant shift of half of the wavelength
\begin{equation}
\begin{aligned}
\langle S_j^x\rangle_{t,\lambda}
=
\sin(2j\theta_\lambda)\,A_j^z
\quad\  
\langle S_j^z\rangle_{t,\lambda}
=
\cos(2j\theta_\lambda)\,A_j^z\,.
\end{aligned}
\end{equation}
This is in agreement with Eqs.~(\ref{Eq:SxSz2}).
Thus directly calculated DMRG results confirm that due to the SOC, the spin vector spirals in the $xz$-plane as one traverses through the sites displaced along $x$ direction.
To extract the dominant wavelengths, $2\pi/k_{\lambda,-}$, in the magnetization profile, we calculated the spin-spin correlations $\left<S^{z}_{10} S^{z}_{10 + l}\right>$ for the system size $L=200$, $D=100$ and without any external field $h$.
To improve the computational efficiency and numerical accuracy, we also fix the total number of particles to half-filling directly in the DMRG algorithm.
The correlations $\left<S^{z}_{10} S^{z}_{10 + l}\right>$ are calculated between the $10^{\text th}$ and $10^{\text th} + l$ site. 
We plot the absolute values of the correlations for $\lambda = \{0.025, 0.05, 0.1, 0.2\}$ in Fig.~\ref{fig:corr_mi3_U10_dim100}.
The correlations seem to decay exponentially with $l$ which is caused by a relatively small bond dimension cut used $D=100$. 
We also present the corresponding fast Fourier transform (FFT) analysis which was used to obtain the dominant wavelengths.
Although the waves in the correlations seem to be very long by eye, the wavelengths are just slightly above the value of $2$. 
The dominant wavelength as a function of $\lambda$ is shown in Fig.~\ref{fig:dft_period_vs_lambda_mu3_U10_paper}. 
We can see that the dominant wavelength is increasing approximately linearly from $2$ (pure AFM ordering) at $\lambda = 0$ as we increase $\lambda$.
We obtained the same numerical results (same wavelengths for the same values of $\lambda$ as before) at another point in the half-filling region, namely $\mu = 2$, $U = 4$, and we expect the same behavior in the whole half-filling region.
The understanding comes from Eq.~(\ref{Eq:HalfFillingKvectors}), particularly,
\begin{equation}
    \text{wavelength}(\lambda)
    =
    \frac{2\pi}{k_{\lambda,-}}
    =
    \frac{2\pi}{\pi-2\arctan{\frac{\lambda}{t}}}
    \,\mathrel{\underset{\lambda\ll t}{\simeq}}\,
    2+\frac{4}{\pi}\frac{\lambda}{t},
\end{equation}
what is remarkably close to the slope in Fig.~\ref{fig:dft_period_vs_lambda_mu3_U10_paper}.


\subsection{Above-half filling}
\begin{figure}
\includegraphics[width=0.48\textwidth]{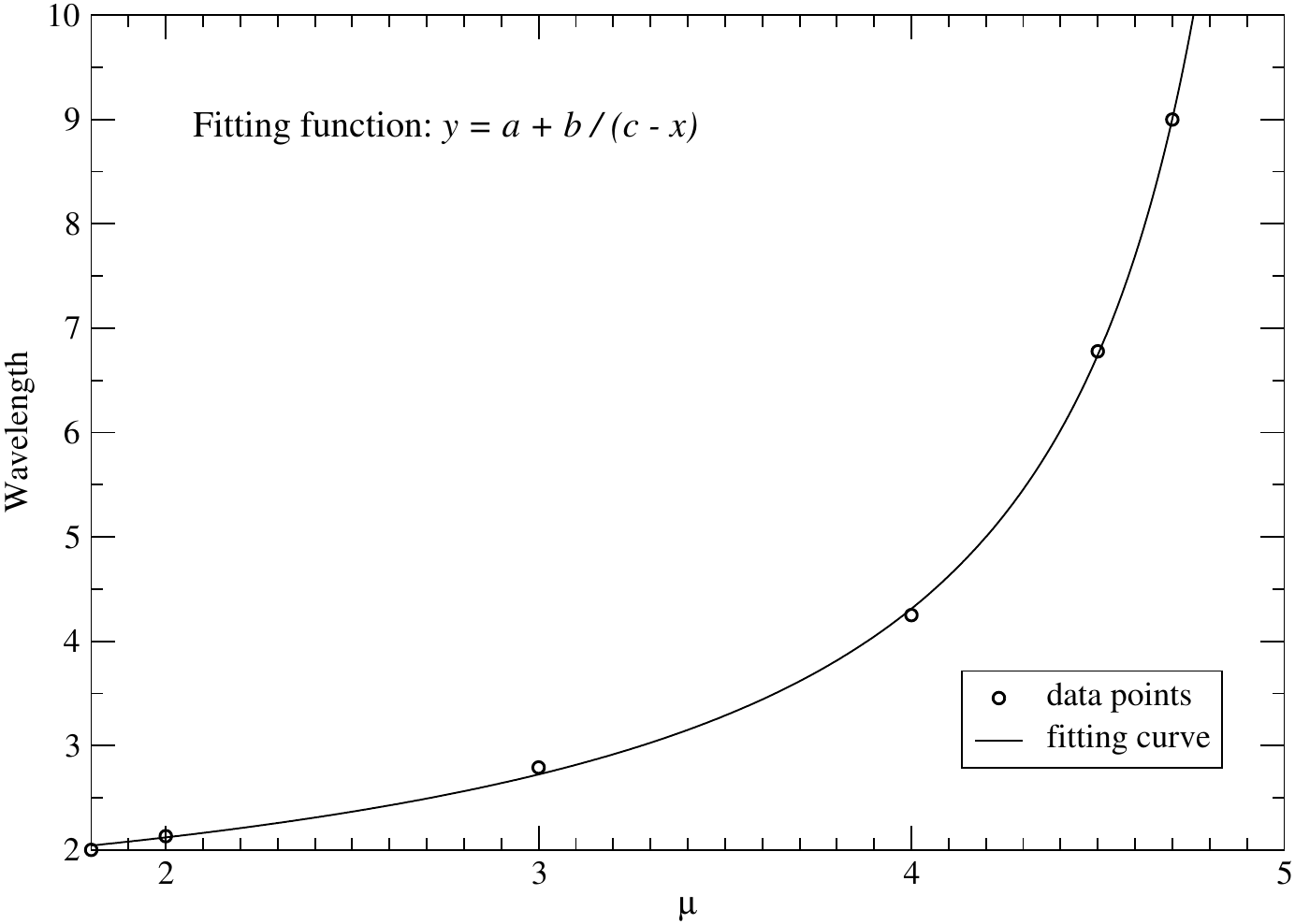}
\caption{
%
SOC-free reference wavelength in the above-half-filled regime. The dominant wavelength of the boundary-pinned $\langle S^z_j\rangle$ profile is shown as a function of chemical potential $\mu$ at $U=3$ and $\lambda=0$. Moving toward the full-band boundary reduces the folded magnetic wave vector and increases the real-space wavelength. The fit $a+b/(c-\mu)$ captures the divergence near the full-filling boundary, with $c\simeq5.2$, close to the expected boundary near $\mu=5$. The data were obtained from open chains with $L=100$--$224$, $D=100$--$200$, and a weak antiparallel boundary field.
}
\label{fig:rpf_u3}
\end{figure}

For $U=3$ and $\lambda=0$, the above-half filling region starts at $\mu \approx 1.8$ and ends at $\mu \approx 5$ (see the phase diagram in Fig.~\ref{fig:phase_diagram_lambda_0}). 
In the half-filling case, $k_0=\pi$ and the corresponding dominant spin modulation wavelength, $2\pi/k_0$, equals 2. 
This wavelength increases once doping $\mu$ reaches the above-half filling region. In Fig.~\ref{fig:rpf_u3} we plot dominant spin modulation wavelength$(\mu)$ obtained from the expectation values of $S^z$ by means of FFT.
Data suggests that the wavelength diverges as $\mu$ approaches the boundary of the full-filling region.
We present the underlying data (the $S^z$ profiles and the corresponding Fourier analysis) in Appendix~\ref{appendix:RPFnoSOC}.\\

\begin{figure}
\includegraphics[width=0.48\textwidth]{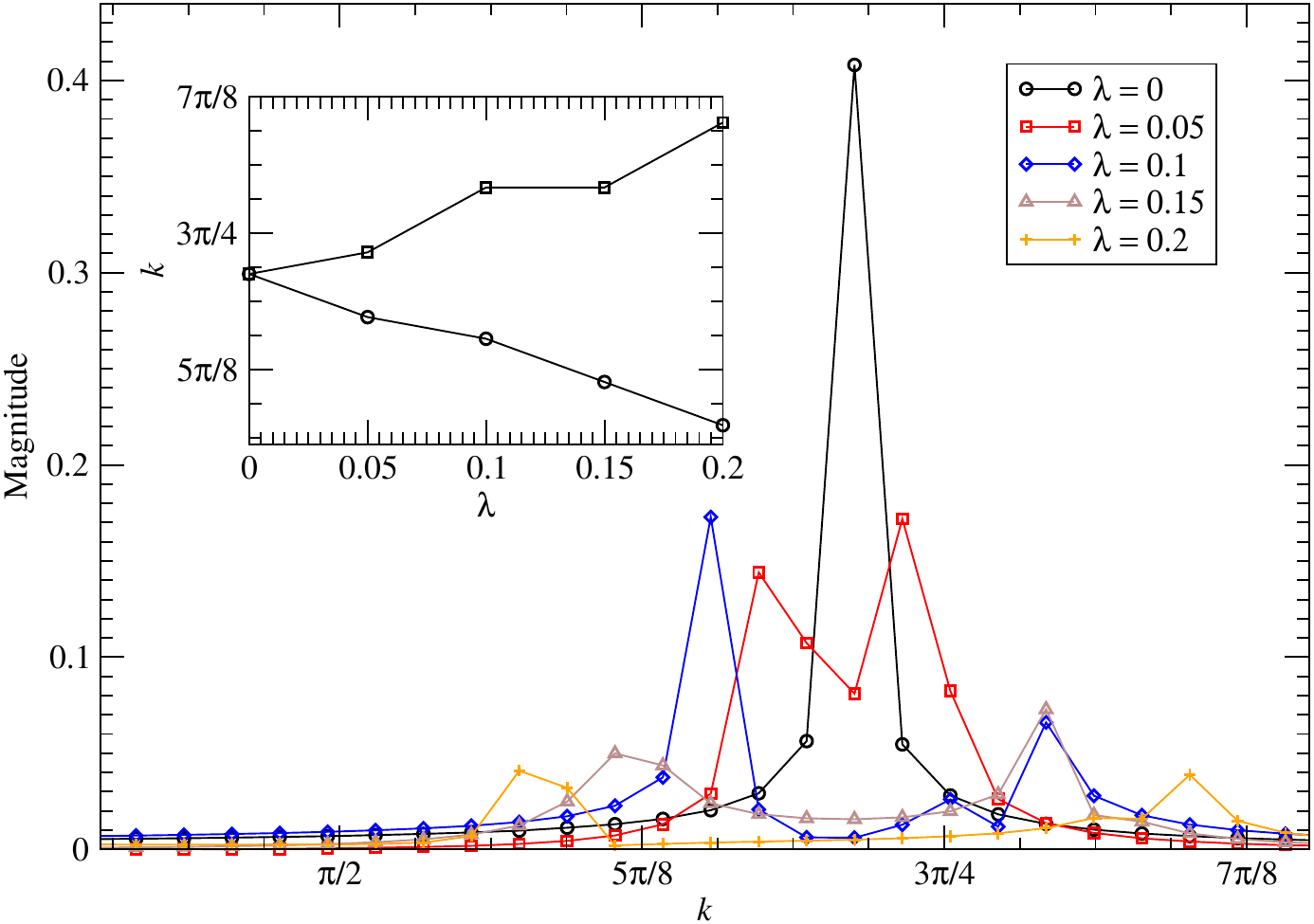}
\caption{
%
SOC-induced sideband splitting of the dominant spin modulation in the above-half-filled regime. The main panel shows the Fourier spectrum of the boundary-pinned $\langle S^z_j\rangle$ profile for $\mu=3$, $U=3$, and several values of $\lambda$. At $\lambda=0$, the spin response has a single dominant wave vector $k_0\simeq2.24$. For $\lambda>0$, this peak splits into two components $k_{\lambda,\pm}=k_0\pm2\arctan(\lambda/t)$, folded into the open-chain Brillouin zone. The inset shows the extracted peak positions as a function of $\lambda$. Parameters: $L=200$, $D=200$, and $t=1$.
}
\label{fig:fft_freq_period_vs_lam_mu3_U3}
\end{figure}

Now we explore the effect of SOC. 
For this, we use system size $L=200$ and bond dimension cut $D=200$.
Figure~\ref{fig:fft_freq_period_vs_lam_mu3_U3} displays what happens 
at $\mu=3$ and $U=3$, where the total number of particles $N=258$, 
with the corresponding filling $f=0.645$.
When $\lambda=0$, we observe only one peak in the dominant wavelength $\approx 2.81$,
corresponding wave-vector $k_0=\frac{2\pi}{2.81}\approx 2.24$.
We notice that the obtained values for $L=200$ and $D=200$ are very 
close to those obtained for $L=102$ and $D=100$ (see Fig.~\ref{fig:fft_w_sz_mu3_U3}) indicating that we do not need to increase the system size $L$ nor the bond dimension $D$ to obtain reliable and representative DMRG results. 

\begin{figure}
\includegraphics[width=0.48\textwidth]{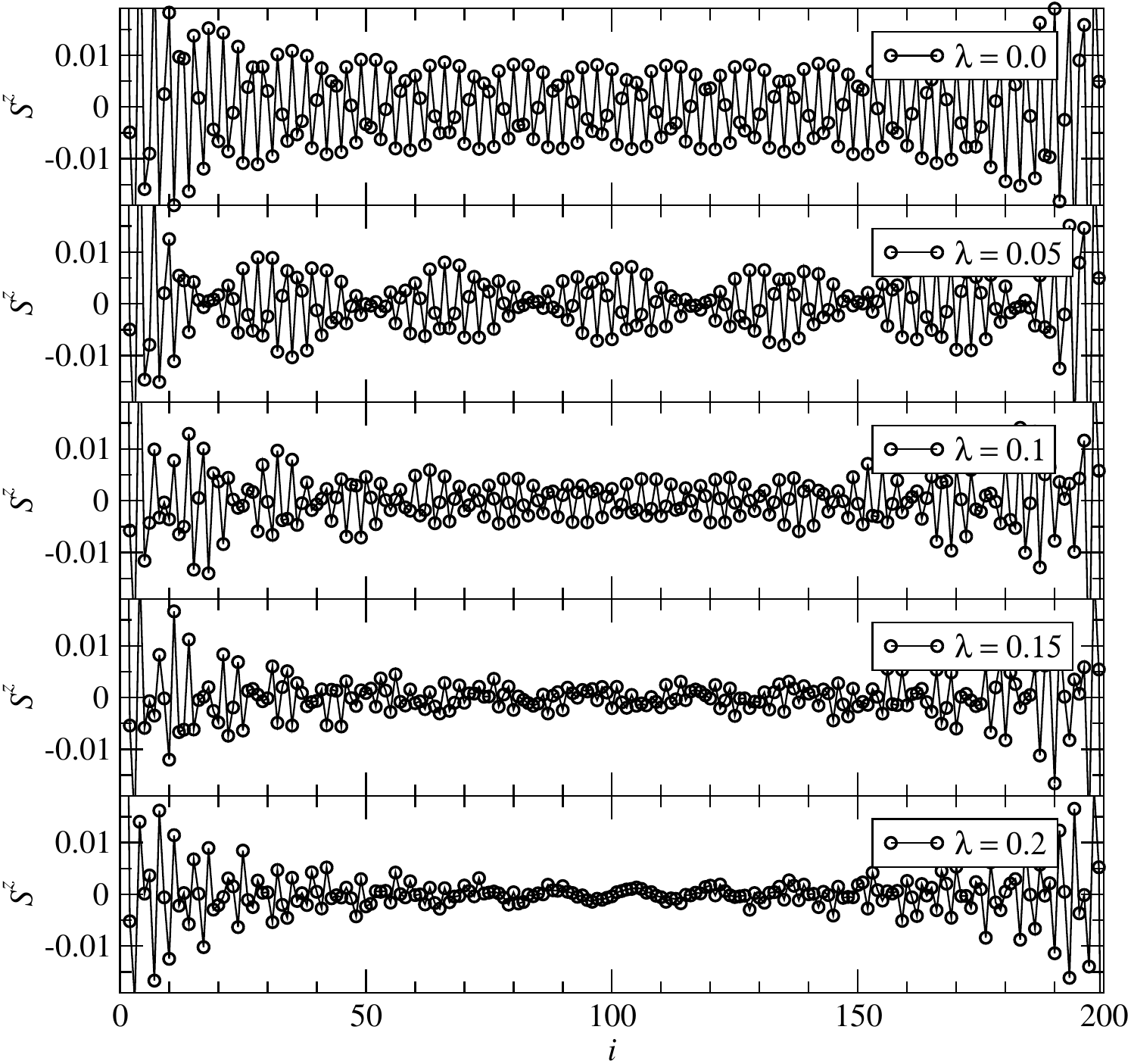}
\caption{
%
Real-space $S^z$ profiles corresponding to the Fourier spectra in Fig.~\ref{fig:fft_freq_period_vs_lam_mu3_U3}. The profiles were obtained at $\mu=3$, $U=3$, and $\lambda=0$, $\lambda=0.05$, $\lambda=0.1$, $\lambda=0.15$, and $\lambda=0.2$. In the absence of SOC, the profile is governed by a single incommensurate Hubbard-chain wave vector. Finite SOC generates two nearby sidebands, producing a beating pattern in real space. The sideband structure is more clearly resolved in momentum space, as shown in Fig.~\ref{fig:fft_freq_period_vs_lam_mu3_U3}. Parameters: $L=200$, $D=200$, and $t=1$.
}
\label{fig:Sz_profiles_mu3_U3_N200_D200_lambdas}
\end{figure}

As $\lambda$ grows the dominant wave-vector peak splits into two, $k_{\lambda,\pm}=k_0\pm 2\arctan(\lambda/t)$, and their separation
\begin{equation}
    \Delta k_\lambda=k_{\lambda,+}-k_{\lambda,-}=4\arctan(\lambda/t)
\end{equation}
grows with increasing $\lambda$ (see the inset in Fig.~\ref{fig:fft_freq_period_vs_lam_mu3_U3}). 
For reference, we present the $S^{z}$ expectation profiles for $\lambda=\{0, 0.05, 0.1, 0.15, 0.2\}$ used in this analysis in Fig.~\ref{fig:Sz_profiles_mu3_U3_N200_D200_lambdas}. 
Although the beating pattern in real space does not allow the two dominant wavelengths to be identified reliably by eye, the corresponding two components are clearly resolved in the FFT spectrum.

\begin{figure}
\includegraphics[width=0.48\textwidth]{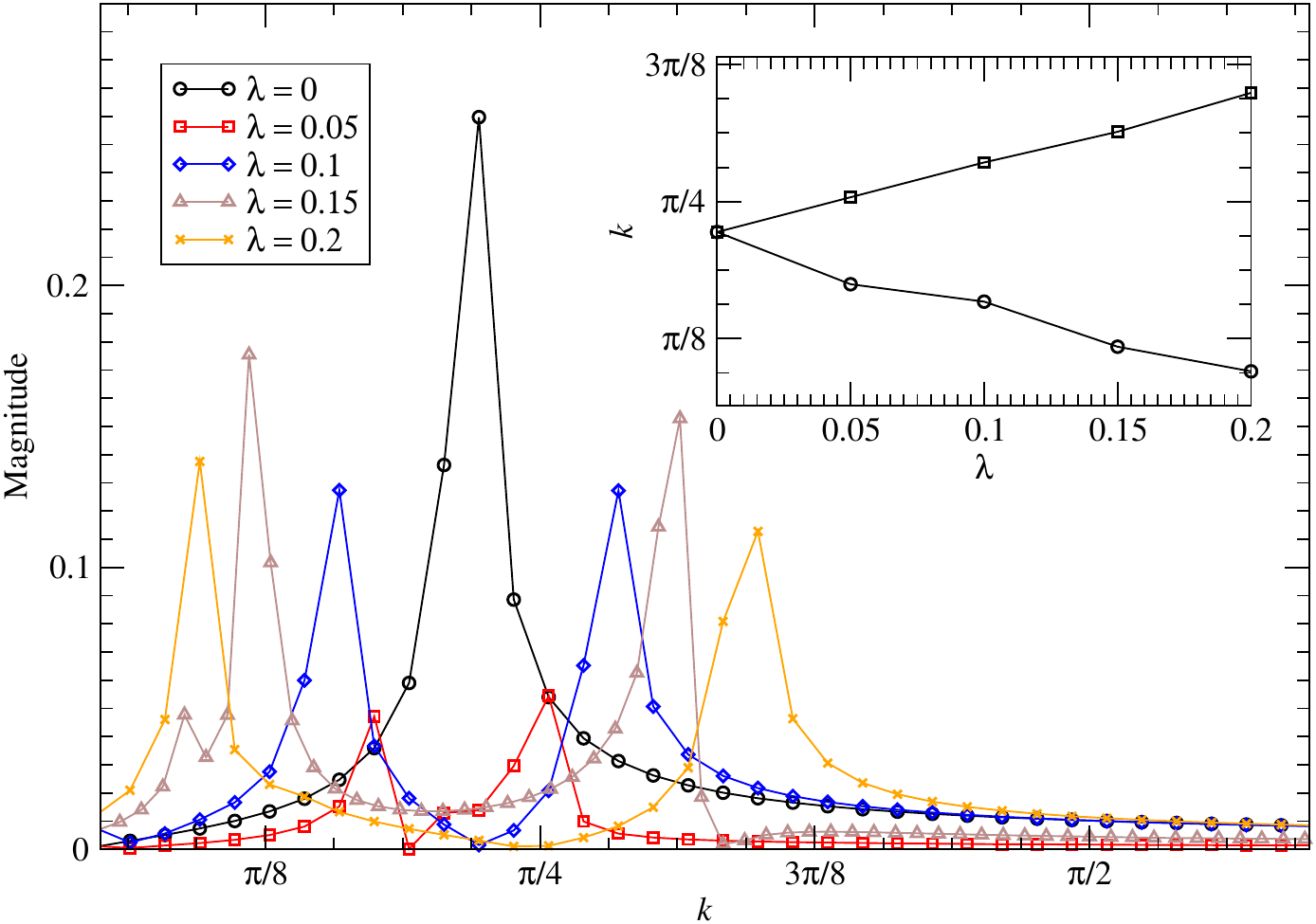}
\caption{
%
SOC-induced sideband splitting closer to the full-filling boundary. The main panel shows the Fourier spectrum of the boundary-pinned $\langle S^z_j\rangle$ profile for $\mu=4.7$, $U=3$, and several values of $\lambda$. At $\lambda=0$, the dominant wave vector is $k_0\simeq0.70$, corresponding to a long-wavelength spin modulation. Finite SOC splits this peak into the two predicted sidebands $k_{\lambda,\pm}=k_0\pm2\arctan(\lambda/t)$, with the extracted peak positions shown in the inset. Parameters: $L=224$, $D=200$, and $t=1$.
}
\label{fig:fft_freq_period_vs_lam_mu4.7_U3}
\end{figure}

We have verified that, for \(\lambda>0\), two dominant wavelengths are also obtained from the 
spin-spin correlations 
\(\left\langle S^{z}_{10} S^{z}_{10+r}\right\rangle\), without the need to use a pinning magnetic field \(h\) at the chain edges. 
This independently confirms the splitting and shifting of the dominant spin modulations in the Rashba-Hubbard chain.
In contrast to the case when $\lambda = 0$, we observe that there is a non-zero $S^{x}$ expectation with some modulation depending on the value of $\lambda$ (not analyzed here). 
To confirm this is a general behavior, we perform a similar analysis for another above-half-filling point, $\mu=4.7$ and $U=3$, see Fig.~\ref{fig:fft_freq_period_vs_lam_mu4.7_U3}.
For $\lambda = 0$, the total number of particles was found to be $N=400$ when $L=224$ (and $D=200$), which gives the total filling $f=N / 2 L \approx 0.89$.
As before, when $\lambda=0$, we observe only one peak in the spin modulation data, 
FFT of the $S^{z}$ expectation values, gives the dominant wavelength $\approx 9.00$ and the corresponding wave-vector $k_0\approx 0.7$, 
see Figs.~\ref{fig:fft_freq_period_vs_lam_mu4.7_U3}~and~\ref{fig:fft_w_sz_mu4.7_U3}. Expectation values of transverse spins $S^{x,y}$ are zero.
As expected, the $S^{z}$ peak splits into two when $\lambda > 0$ and the distance between these two peaks increases as $\lambda$ increases, see the inset of Fig.~\ref{fig:fft_freq_period_vs_lam_mu4.7_U3}. Also we observe that there is a non-zero $S^{x}$ expectation with a modulation depending on the value of $\lambda$, but we did not analyze that in greater details. 

As expected from the particle-hole symmetry the same features shall be present in the under-half filling region. We verify that and results are summarized in Appendix~\ref{App:under-half filling}.

\section{Conclusions and outlook}\label{sec:conclusions}

We have used DMRG to study the repulsive one-dimensional Rashba-Hubbard chain in the regime where uniform spin-orbit coupling is exactly removable from the open-chain Hamiltonian. This property makes the model a useful benchmark rather than a conventional search problem for new SOC-induced phases. For open boundary conditions, the site-dependent spin rotation maps the model with hopping $t$ and SOC strength $\lambda$ onto the ordinary Hubbard chain with renormalized hopping $t_\lambda=\sqrt{t^2+\lambda^2}$. As a result, the filling structure and the charge-sector diagnostics follow the corresponding SOC-free Hubbard model. In the weak-SOC regime this effect enters through the bandwidth renormalization and is therefore quadratic in $\lambda/t$.

The central result is that the same SOC field has a qualitatively sharper effect in the laboratory-frame spin sector. Although the Hamiltonian can be mapped to the ordinary Hubbard form, the spin operators are locally rotated along the chain. Therefore SOC acts as a spin-correlation wave-vector transducer: a dominant Hubbard-chain magnetic wave vector $k_0$ is shifted into sidebands at $k_0\pm k_{\rm so}$, where $k_{\rm so}=2\arctan(\lambda/t)$, with the result folded into the open-chain Brillouin zone. This produces a response that is linear in $\lambda/t$ at weak SOC, in contrast to the quadratic response of charge and energy diagnostics.

DMRG resolves this sideband mechanism directly in real space and across different filling regimes. At half filling, the ordinary repulsive Hubbard chain has dominant staggered spin correlations with $k_0=\pi$. The two shifted components $\pi\pm k_{\rm so}$ are equivalent after folding, and the spin response appears as a single SOC-induced spiral modulation with folded wave vector $k=\pi-k_{\rm so}$. Away from half filling, the Hubbard-chain spin response is already incommensurate. The same SOC rotation then produces two distinct shifted components rather than one folded component, giving a real-space beating pattern and a clear two-peak structure in the Fourier spectra.

The role of DMRG in this work is therefore not merely to reproduce an exact transformation. It is to test how the transformation is manifested in finite open chains, where boundary pinning, finite bond dimension, finite Fourier resolution, and filling-dependent incommensurability all affect the observable spin profiles and spin-spin correlations. In this sense, the present calculation supplies a controlled real-space and filling-resolved benchmark for tensor-network studies of SOC-coupled correlated chains.

This benchmark also clarifies what uniform one-dimensional SOC alone can and cannot do. In the open single-band Hubbard chain, it does not generate a new bulk phase diagram. Its robust fingerprint is instead the predictable displacement and splitting of spin-correlation weight. This distinction is important for interpreting more complex SOC-Hubbard systems, where the exact open-chain mapping no longer applies. Natural extensions include periodic chains, where SOC becomes a spin-dependent boundary twist; ladders and multiorbital chains, where the SOC field generally cannot be removed by a single local rotation; and proximitized or Zeeman-coupled wires, where SOC can participate directly in helical and topological superconducting physics.

Future tensor-network studies can use the present result as a reference point. Deviations from the sideband law $k_0\to k_0\pm2\arctan(\lambda/t)$ would then signal genuinely nontrivial SOC-interaction physics beyond the removable single-band limit. Such deviations may arise from non-Abelian SOC textures, longer-range interactions, disorder, orbital degrees of freedom, superconducting proximity coupling, or higher-dimensional geometries. The present work thus identifies the exactly constrained baseline from which those less trivial correlated spin-orbit systems can be systematically analyzed.

\bigskip

\begin{acknowledgments}
We thank Juraj Hasik and Manuel Schneider for valuable discussions.
Funded by the EU NextGenerationEU through the Recovery and Resilience Plan for Slovakia under the project No. 09I03-03-V04-00682; by the Slovak Research and Development Agency, grant No.~APVV-24-0134 and No.~APVV-24-0091; by Vedeck\'{a} Grantov\'{a} Agent\'{u}ra M\v{S}VVaM SR and SAV through the grant VEGA No.~2/0152/26; by project IM-2021-26 (SUPERSPIN) funded by the Slovak Academy of Sciences via the programme IMPULZ 2021; and
by National Science and Technology Council of Taiwan, grant No.~114-2112-M-006-034-MY3.
C.-M. C. acknowledges support by MOST (114-2628-M-A49-007-MY3, and 115-2124-M-007-015).
\end{acknowledgments}

\bibliographystyle{apsrev4-2}
\bibliography{references}

\clearpage

\appendix

\section{Outer boundaries of the phase diagram} \label{appendix:bound}

In the case of the non-interacting model
\begin{equation}
    {\cal H} = H_0 +  H_{\rm soc},
\end{equation}
one can directly obtain energy spectrum
\begin{equation}
    e_{k\pm} = -\mu -2\sqrt{t^2 + \lambda^2}\cos\left( k\pm\arctan\left(-\frac\lambda t\right)\right).
\end{equation}
If we set $\lambda=0$ then in order to populate the ground state with particles the chemical potential $\mu$ has to obey $\mu>-2|t|$.
This is the reason why the boundary between empty and non-empty states is 
\begin{equation}
\mu_b = -2|t|.
\end{equation}

We can also estimate a shift of this boundary by switching on the spin-orbit hopping $\lambda$. 
Up to the second order in $\lambda$ the new boundary is 
\begin{equation}
    \mu_b(\lambda) = -2|t| - \frac{\lambda^2}t.
\end{equation}

Situation on the opposite side between the fully occupied and partially occupied ground states is analogous, and follows from particle-hole symmetry, see Eq.~(\ref{Eq:PHsymmetry}). Transition from the non-fully-occupied to fully occupied state will appear roughly at


\begin{equation}
\mu_b(\lambda) = 2|t|+\frac{\lambda^2}t+U.
\end{equation}

\section{Above-half filling without SOC: $S^z$ profiles and Fourier analysis} \label{appendix:RPFnoSOC}

\begin{figure}
\includegraphics[width=0.48\textwidth]{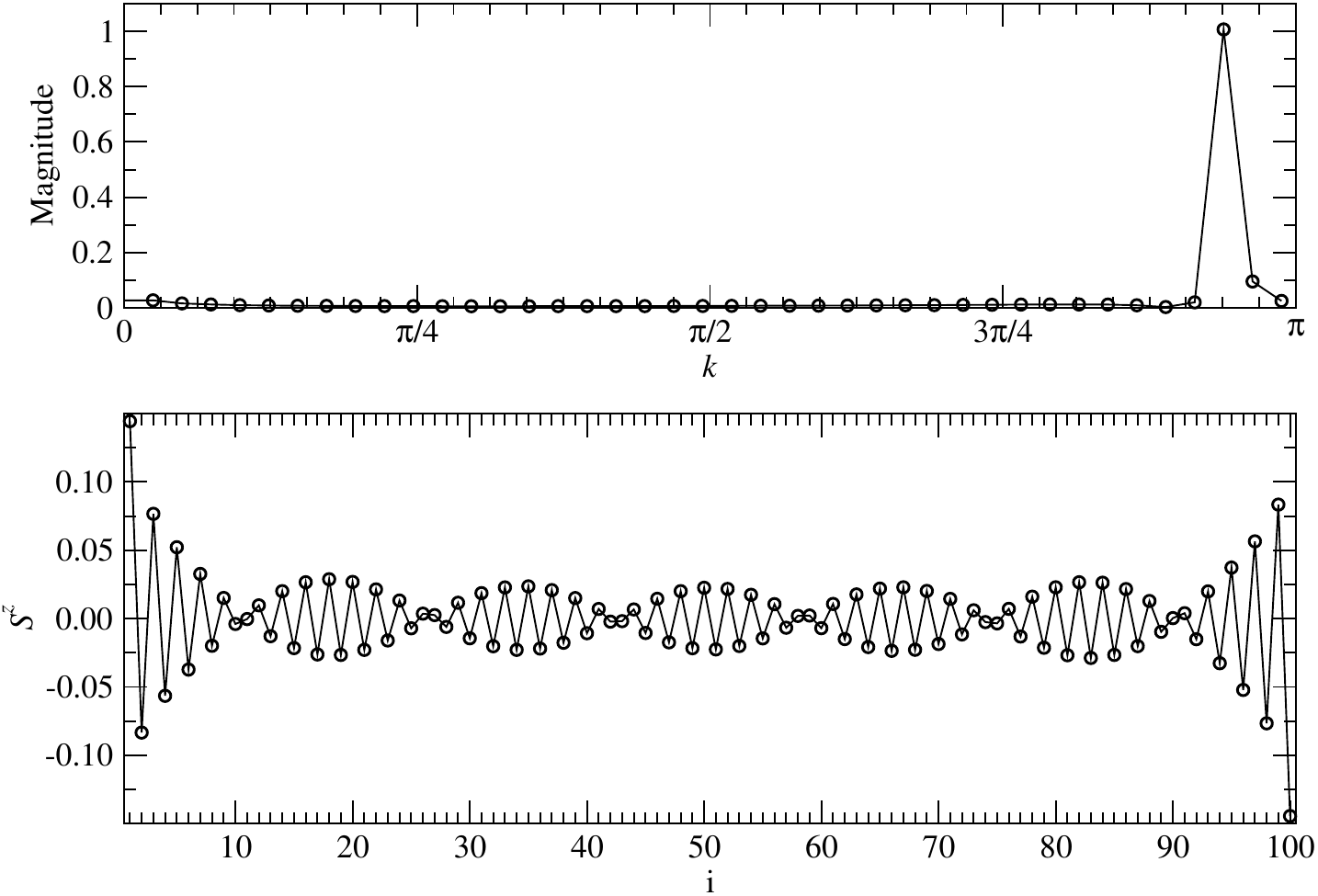}
\caption{
%
SOC-free boundary-pinned spin profile and Fourier spectrum for the above-half-filled chain at $\mu=2$ and $U=3$. The dominant Fourier peak gives a wavelength $\ell\simeq2.13$, slightly longer than the half-filled staggered value $\ell=2$. This data point provides the first SOC-free reference wave vector used to construct the filling-dependent baseline in Fig.~\ref{fig:rpf_u3}. Parameters: $\lambda=0$, $L=100$, $D=100$, and $t=1$. Boundary sites were excluded from the Fourier analysis to reduce edge effects.
}
\label{fig:fft_w_sz_mu2_U3}
\end{figure}

\begin{figure}
\includegraphics[width=0.48\textwidth]{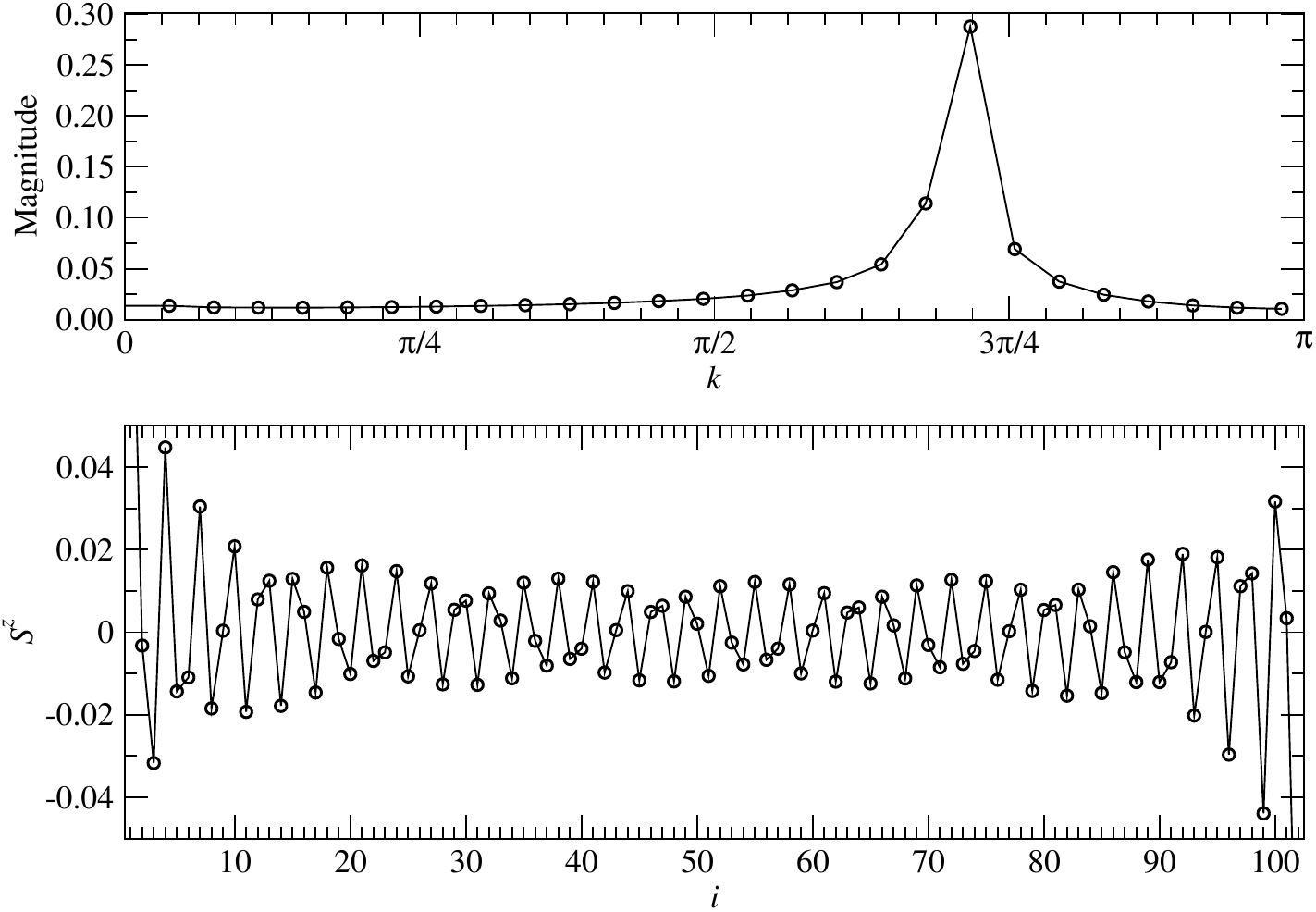}
\caption{
%
SOC-free boundary-pinned spin profile and Fourier spectrum for $\mu=3$ and $U=3$ in the above-half-filled regime. The dominant wavelength is $\ell\simeq2.79$. The apparent multiple-wave structure in real space results from the non-integer wavelength of a single dominant incommensurate component. Parameters: $\lambda=0$, $L=102$, $D=100$, and $t=1$. Boundary regions were excluded from the Fourier analysis.
}
\label{fig:fft_w_sz_mu3_U3}
\end{figure}

\begin{figure}
\includegraphics[width=0.48\textwidth]{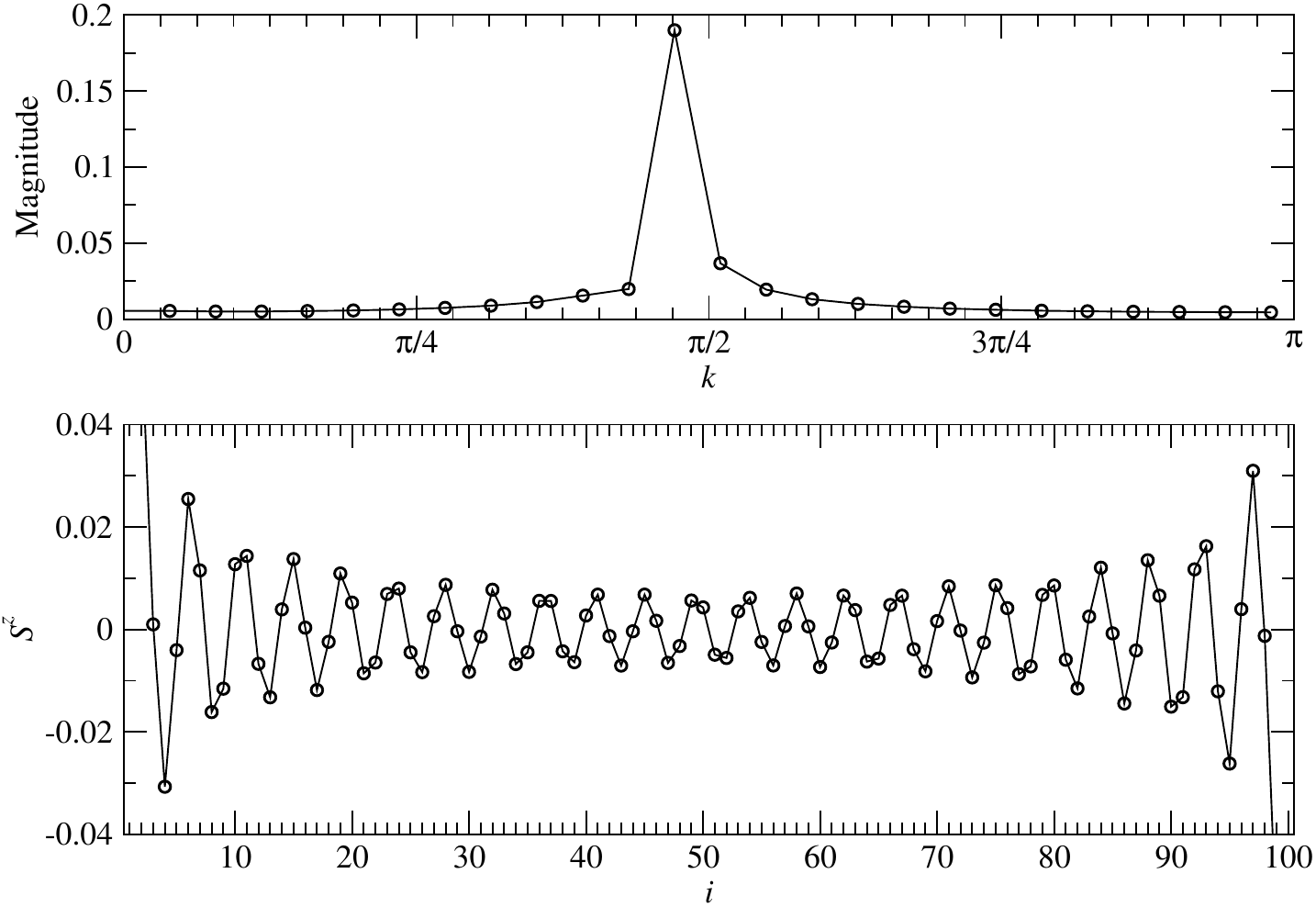}
\caption{
%
SOC-free boundary-pinned spin profile and Fourier spectrum for $\mu=4$ and $U=3$ in the above-half-filled regime. The dominant wavelength increases to $\ell\simeq4.25$, reflecting the continued decrease of the folded magnetic wave vector as the system approaches the full-filling boundary. Parameters: $\lambda=0$, $L=100$, $D=100$, and $t=1$.
}
\label{fig:fft_w_sz_mu4_U3}
\end{figure}

\begin{figure}
\includegraphics[width=0.48\textwidth]{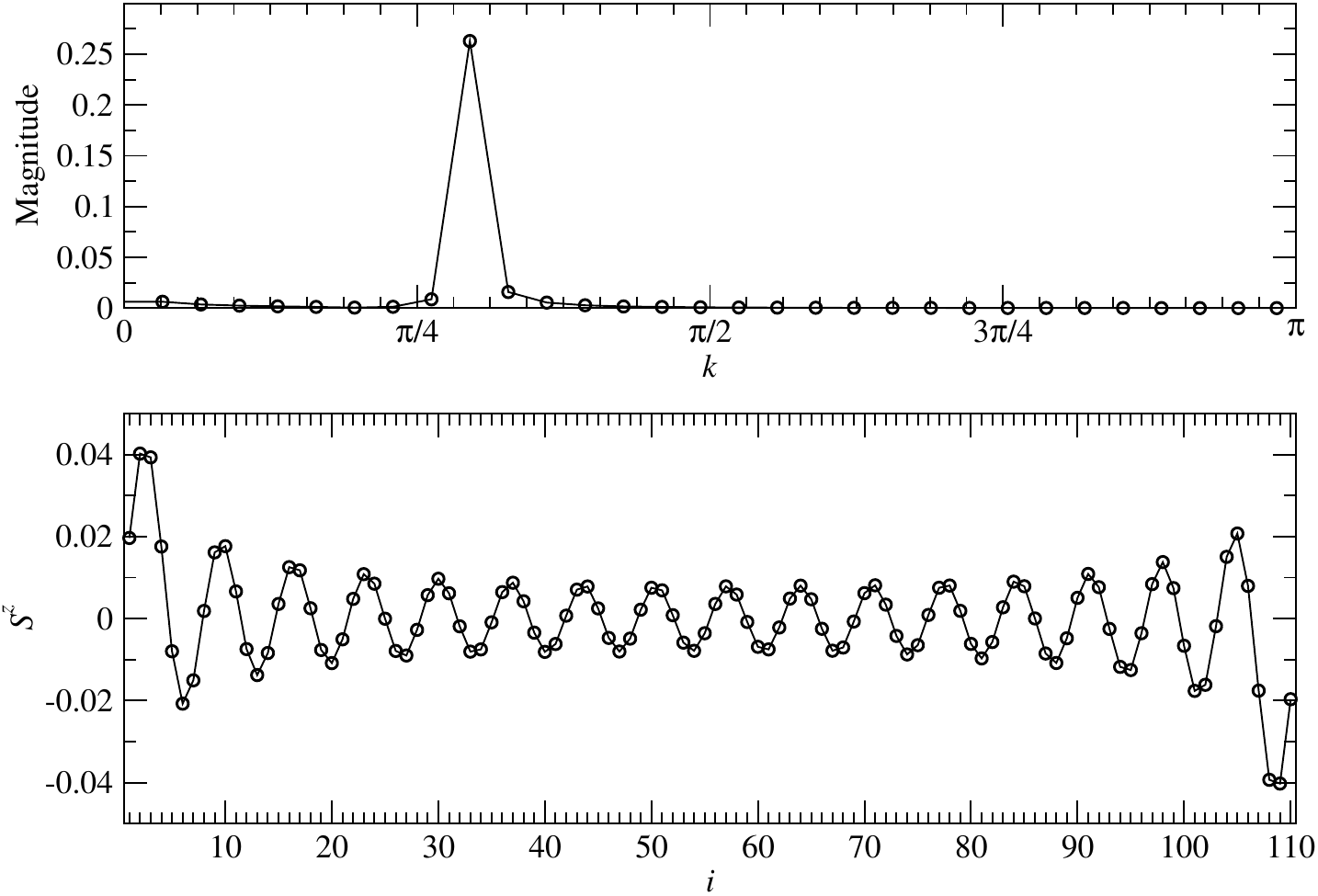}
\caption{
%
SOC-free boundary-pinned spin profile and Fourier spectrum for $\mu=4.5$ and $U=3$. The dominant wavelength is $\ell\simeq6.78$, continuing the trend shown in Fig.~\ref{fig:rpf_u3}: the spin modulation becomes longer ranged in real space as the filling approaches the full band. Parameters: $\lambda=0$, $L=110$, $D=100$, and $t=1$.
}
\label{fig:fft_w_sz_mu4.5_U3}
\end{figure}

\begin{figure}
\includegraphics[width=0.48\textwidth]{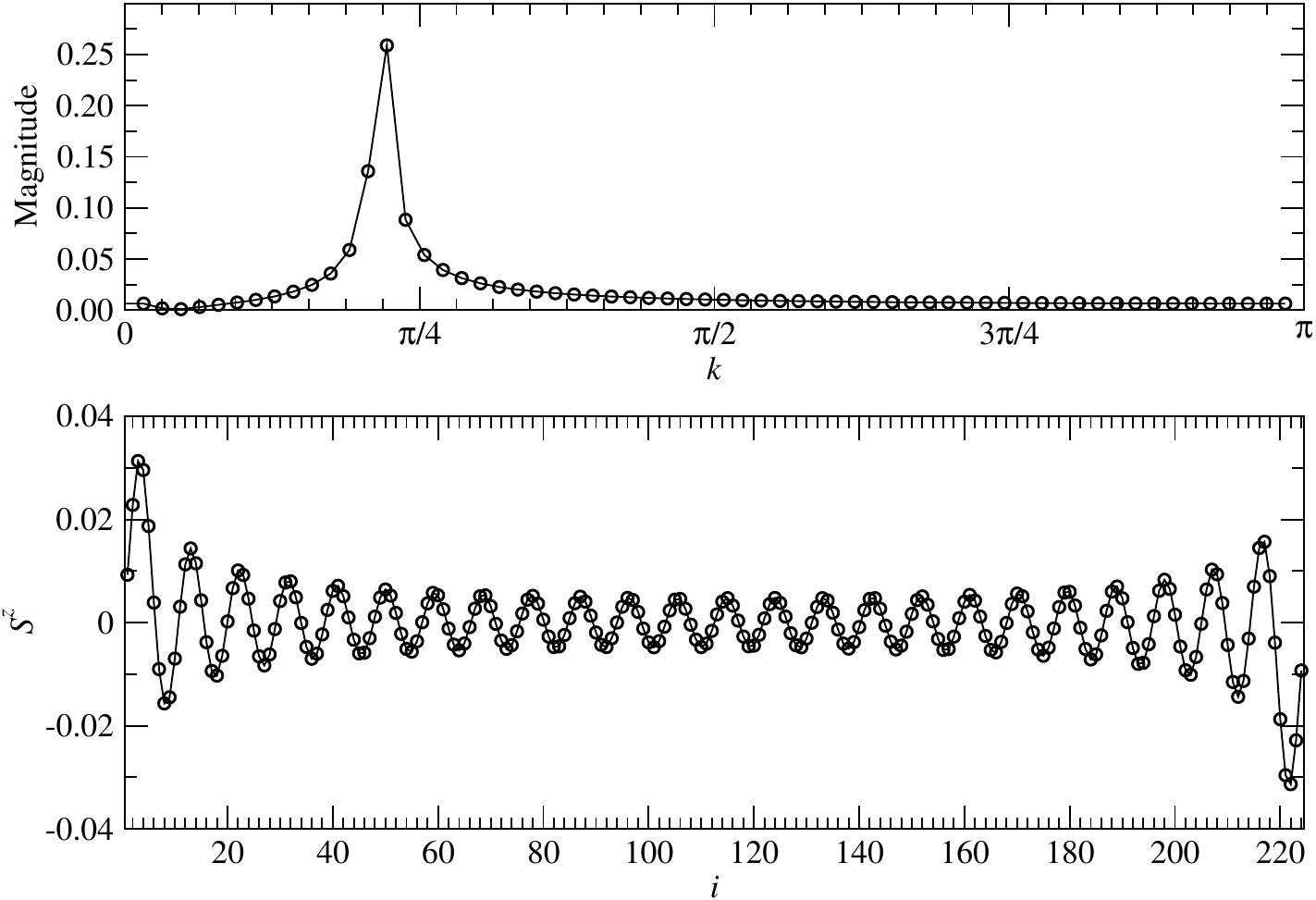}
\caption{
%
SOC-free boundary-pinned spin profile and Fourier spectrum close to the full-filling boundary, at $\mu=4.7$ and $U=3$. The dominant wavelength is $\ell\simeq9.00$, corresponding to the small baseline wave vector $k_0\simeq0.70$ used in the SOC sideband analysis of Fig.~\ref{fig:fft_freq_period_vs_lam_mu4.7_U3}. Parameters: $\lambda=0$, $L=224$, $D=200$, and $t=1$.
}
\label{fig:fft_w_sz_mu4.7_U3}
\end{figure}

Let us start with $\mu = 2$, $U = 3$, see Fig.~\ref{fig:fft_w_sz_mu2_U3}. 
We used system size $L=100$ and found the dominant wavelength to be $\approx 2.13$. 
The visible modulation of $S^{z}$ expectation values is caused by the dominant wavelength 
that is slightly above $2$.
Expectation values of $S^{x}$ are zero. 
The $S^{z}$ expectation and Fourier analysis for $\mu = 3$, $U = 3$ is in Fig.~\ref{fig:fft_w_sz_mu3_U3}. 
The dominant wavelength was found to be $\approx 2.79$.
One might notice three ``waves'' going through each other in the $S^{z}$ expectation; however, this effect is caused just by having a dominant wavelength being a non-integer value close to three. 
The $S^{z}$ expectation and Fourier analysis for $\mu = 4$, $U = 3$ is in Fig.~\ref{fig:fft_w_sz_mu4_U3}. 
The dominant wavelength was found to be $\approx 4.25$.
%
%
The $S^{z}$ expectation of Fourier analysis for $\mu = 4.5$, $U = 3$ is in Fig.~\ref{fig:fft_w_sz_mu4.5_U3}. 
The dominant wavelength was found to be $\approx 6.78$.
The $S^{z}$ expectation of Fourier analysis for $\mu = 4.7$, $U = 3$ is in Fig.~\ref{fig:fft_w_sz_mu4.7_U3}. 
The dominant wavelength was found to be $\approx 9.00$.

\section{Entanglement entropy in the partially filled regime} \label{appendix:Entropy}

\begin{figure}
\includegraphics[width=0.48\textwidth]{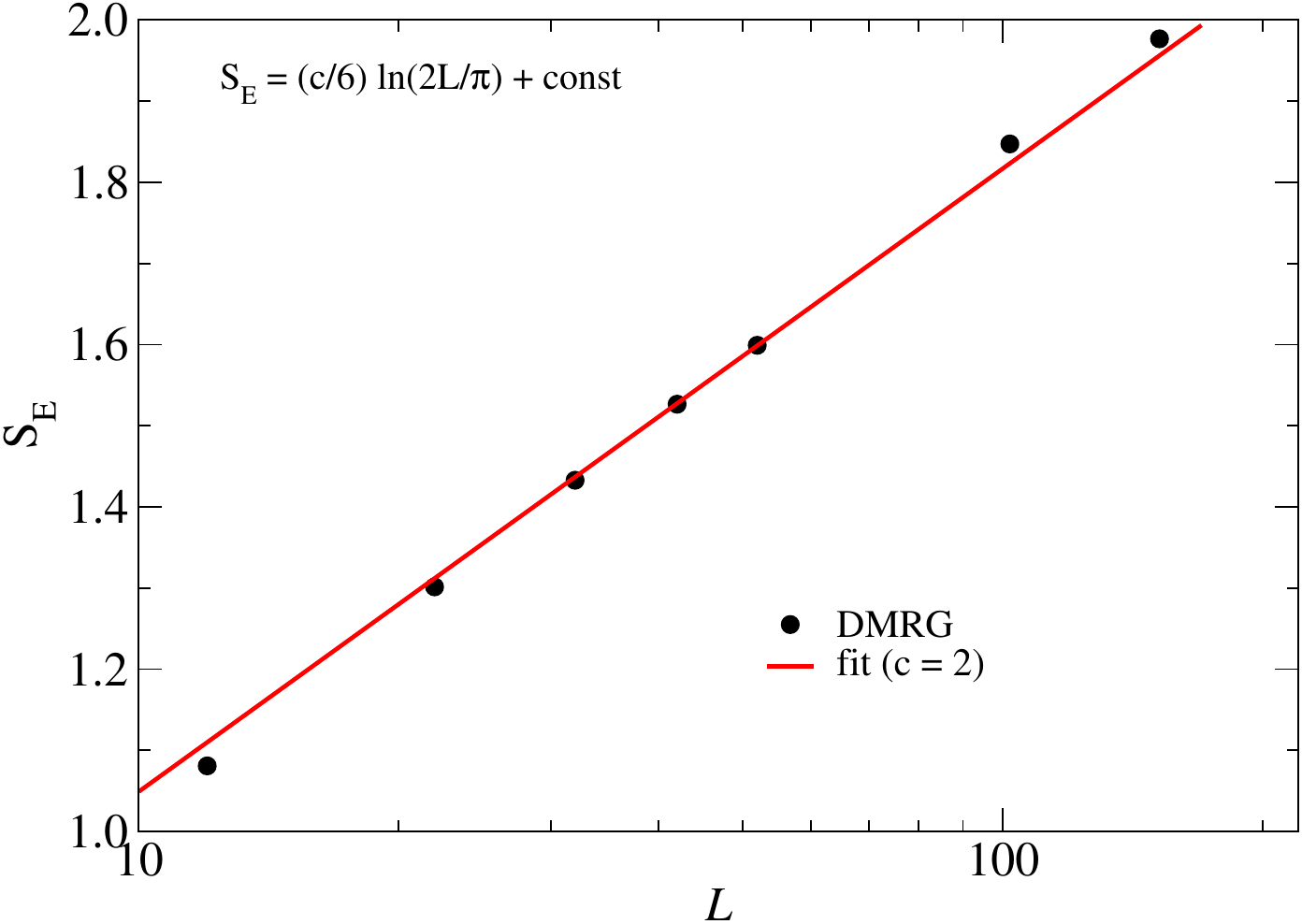}
\caption{
Entanglement entropy at the center of the system $L/2$ as a function of system size $L$, for $U=4$, $\mu=-1$, and bond dimension up to $D=300$. Semilogarithmic scale. Entanglement entropy grows logarithmically, which is a hallmark of the critical Luttinger liquid (LL) phase. Central charge obtained from fitting the slope of the function in semi-logarithmic scale is $c=2$, which corresponds to two independent modes. Charge and spin modes for the non-SOC case and their combination for the SOC case.
}
\label{fig:vNe}
\end{figure}

Entanglement entropy $S_E$ at the center of the system $L/2$ as a function of the system size $L$ has been calculated in Fig.~\ref{fig:vNe}. It can be clearly seen that it grows logarithmically, which is a hallmark of the critical gapless phase~\cite{calabrese2004entanglement}, in this case, the Luttinger liquid (LL) phase with central charge $c=2$
\begin{equation}
S_E(L) = \frac c6\ln\left(\frac{2L}\pi\right)+\dots.
\end{equation}
Because of locality and unitarity of the transformation connecting systems with and without spin-orbit coupling~\cite{kaplan1983single}, entanglement entropy is not changed when applying the transformation. It means that in both cases the system is in the Luttinger liquid phase~\cite{giamarchi2003quantum} with two independent modes, consistent with the logarithmic growth of the entanglement entropy and central charge $c=2$. In the case without SOC, modes correspond to charge and spin modes. In the case of SOC, it is a combination of spin and charge degrees of freedom.

\section{Under-half filling}\label{App:under-half filling}

\begin{figure}
\includegraphics[width=0.48\textwidth]{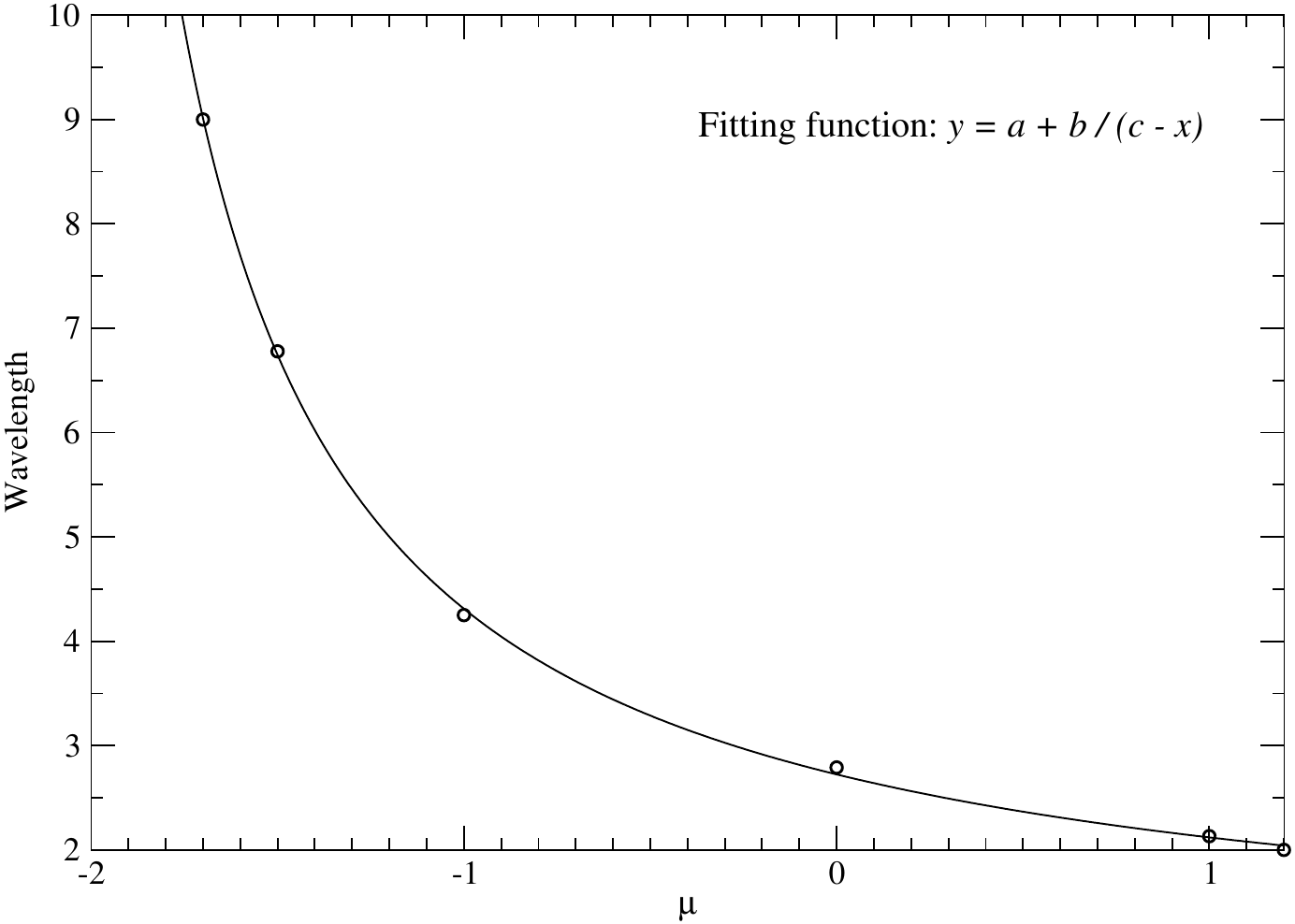}
\caption{
%
SOC-free reference wavelength in the under-half-filled regime. The dominant wavelength of the boundary-pinned $\langle S^z_j\rangle$ profile is shown as a function of chemical potential $\mu$ at $U=3$ and $\lambda=0$. The wavelength increases as $\mu$ approaches the empty-chain boundary and is fitted by $a+b/(\mu-c)$, with the divergence located near $c\simeq-2.2$, close to the expected empty-chain boundary $\mu=-2$. The result mirrors the above-half-filled data in Fig.~\ref{fig:rpf_u3}, as expected from particle-hole symmetry. The data were obtained from open chains with $L=100$--$224$, $D=100$--$200$, and a weak antiparallel boundary field.
}
\label{fig:lpf_u3}
\end{figure}

For $U=3$, without SOC the under-half filling region starts at $\mu \approx -2$ and ends at $\mu \approx 1.2$ (see the phase diagram in Fig.~\ref{fig:phase_diagram_lambda_0}). 
The wavelength increases in the under-half filling region as we decrease $\mu$. 
It seems the wavelength is increasing to infinity as we approach the boundary of the empty filling region.
We plot the wavelength obtained from $S^z$ profile by performing FFT in Fig.~\ref{fig:lpf_u3}.
When comparing Fig.~\ref{fig:lpf_u3} with Fig.~\ref{fig:rpf_u3}, one can notice there is a mirror symmetry. 
In fact, we obtained exactly the same wavelengths and $S^z$ profiles at $\mu = \{1, 0, -1, -1.5, -1.7\}$ as at $\mu = \{2, 3, 4, 4.5, 4.7\}$, respectively.
This behavior can be explained by the particle-hole symmetry of the Hamiltonian.
When for example $\mu=0$ and $L=200$, we observed the total number of particles $N=142$ and wavelength $\approx 2.81$ (for $D=200$).
The corresponding point in the above-half filling region is $\mu=3$, where we found $N=258$ and wavelength $\approx 2.81$.
This means, we have $2 \times L - N = 142$ holes, which is the same as the number of particles in the under-half filling region at $\mu=0$. 
Similarly, when $\mu=-1.7$ and $L=224$, we obtained $N=48$ and wavelength $\approx 9.00$ (for $D=200$).
The corresponding point in the above-half filling region is $\mu=4.7$, where we found $N=400$ and wavelength $\approx 9$.
This means, we have $2 \times L - N = 48$ holes, which is the same as the number of particles in the under-half filling region at $\mu=-1.7$.
Thus, the holes and particles are simply exchanged and the same behavior is reproduced. 
Due to the particle-hole symmetry, there is the same behavior as in the above-half filling region when $\lambda > 0$. 
This means there is the same splitting of the wavelengths as observed at $\mu=3$ (Fig.~\ref{fig:fft_freq_period_vs_lam_mu3_U3}) and $\mu=4.7$ (Fig.~\ref{fig:fft_freq_period_vs_lam_mu4.7_U3}) for the points $\mu = 0$ and $\mu = -1.7$, respectively. 

\end{document}